\documentclass[11pt,a4paper,notitlepage] {article}
\usepackage{amsfonts}
\usepackage{amstext}
\usepackage{amscd}

\def\a{\`a }
\def\e{\`e }

\def\a{\`a }

\def\c{\mathcal}

\def\n{\nabla}
\def\ep{\epsilon}
\def\pa{\partial}
\def\we{\wedge }
\def\phi{\varphi}
\def\be{\begin{equation}}
\def\ee{\end{equation}}
\def\ba{\begin{eqnarray}}
\def\ea{\end{eqnarray}}
\def\baa{\begin{eqnarray*}}
\def\eaa{\end{eqnarray*}}
\def\bc{\begin{center}}
\def\ec{\end{center}}
\def\R{I \kern-.36em R}
\def\E{I \kern-.36em E}
\def\F{I \kern-.36em F}
\newtheorem{definition}{Definition}

\begin{document}

\title{The First Law of Isolated Horizons via Noether Theorem}

\author{G. Allemandi\footnote{E-mail: allemandi@dm.unito.it}, M. Francaviglia\footnote{E-mail:
francaviglia@dm.unito.it}, M. Raiteri\footnote{E-mail: raiteri@dm.unito.it}\\
Dipartimento di Matematica, Universit\a di Torino\\
Via Carlo Alberto 10; 10123 Torino, Italy}

\date{}

\maketitle

\begin{abstract}

A general recipe  proposed elsewhere to define,  via Noether theorem, the variation of energy for
a natural field theory is applied to Einstein-Maxwell theory. 
The electromagnetic field is analysed in the geometric framework of natural bundles.
Einstein-Maxwell theory turns then out to be natural  rather than  gauge-natural. As a
consequence of this assumption  a correction term \a la Regge-Teitelboim  is
needed to define the variation of energy, also for the pure electromagnetic part of the
Einstein-Maxwell Lagrangian. Integrability conditions for the variational equation which defines
the variation of energy are analysed in relation with boundary conditions on physical data. As an
application the first law of thermodynamics for rigidly rotating horizons is obtained.
\end{abstract}

\section{\large Introduction}
In a previous paper \cite{raiteri} a recipe to define, via Noether theorem,
the variation of Noether conserved quantities in natural field theories was proposed. The main
result of the theory there developed was to describe, for vacuum General Relativity, the
quasilocal stress-energy content of a spatially bounded region of spacetime  with non-orthogonal
boundaries. \\
The purpose of the present paper is to generalise the  recipe of \cite{raiteri} to
Einstein-Maxwell theory in order to calculate the Noether quasilocal conserved quantities, still
remaining
 in the general case of spacetime regions with non-orthogonal boundaries. As a remarkable
application we define a first principle of thermodynamics for spacetimes admitting an
\textit{isolated horizon} (in the sense of \cite{Ashtekar1}).\\

\noindent Using a geometric approach to field theories in the framework of natural
theories, it is possible to associate to each infinitesimal symmetry $\xi$ on the base
manifold $M$ (spacetime of dimension $m$) a conserved quantity $Q(\xi)$. This Noether charge
is obtained by integrating the Noether current on a compact $(m-1)$ submanifold $\Sigma$
of $M$ with boundary $\pa \Sigma$ and it can be decomposed into a bulk term $Q_{bulk}
(\xi)$ and a surface term $Q_{surface} (\xi)$. The bulk term is obtained by integrating on
$\Sigma$ the \textit{reduced Noether current}, which is a $(m-1)$-form vanishing on shell (i.e.
along solutions). The surface term is the integral over $\pa \Sigma$ of the \emph{superpotential}
related to the symmetry generated by $\xi$ which in turn, is a $(m-2)$-form algorithmically
calculated starting from the Lagrangian of the theory \cite{Lagrange}. To obtain physically
expected values, in analogy with the original prescription suggested by Regge-Teitelboim in
\cite{RT} (when dealing with the analysis of Hamiltonian boundary terms),  we have to correct the
Noether conserved quantities by suitably adding boundary terms to
$Q(\xi)$; this approach was extensively developed for natural and gauge-natural theories in a
global and geometric framework, called \textit{covariant ADM formalism}. \cite{CADM}.\\
Following the same idea which leads to the covariant ADM formalism, we can proceed one step
further by defining  the \textit{ variation} of the corrected conserved quantities through the
addition of  a further suitable boundary term
$\tau$ (which is still present in the covariant ADM definition of the variation of the conserved
quantity), such that
$\delta Q(\xi)$ turns out to be a pure
 (off-shell) bulk term. As a particular case, if we
consider a Cauchy surface $\Sigma$ in spacetime and a vector field $\xi$ transverse to
$\Sigma$ the variation of the Hamiltonian can be defined as the variation
of the Noether charge associated to $\xi$. In this situation the vector $\xi$ is identified
with the generator of time flow, meaning that the  parameter "$t$" generating the flow
of $\xi$ itself is identified with time. This technique allows us to obtain the Hamiltonian
equations of motion and to define the symplectic structure of the theory together with its phase
space. Variation of energy is simply defined as the on-shell value of the variation of the 
Hamiltonian. A problem which arises is then to formally integrate, if possible, the
variation of the conserved quantities to obtain the explicit expression for the Hamiltonian and
for the energy. This problem  is related to the choice of suitable
boundary conditions as well as to the choice of a background solution. The chosen background
 can be interpreted as a reference solution in the one-parameter
family of solutions satisfying the same boundary conditions 
(namely the one-parameter family along which variations are performed). \\ 
We remark that the final formula for the variation of  the Hamiltonian is independent on 
divergence terms which might be added to the Lagrangian. This is a mathematical and
physical advantage since we do no longer have to take care about surface terms in the action
functional.  The expression of the variation of the Hamiltonian is unique for each element of the
equivalence class 
$[L]$ formed by all the Lagrangians differing  one from the other only for divergence terms.
Moreover, the variation of the Hamiltonian reproduces the correct equations of motion in the
 whole phase space of the theory and not only in a phase space restricted by  boundary
conditions such as, for example, a suitable fall-off requirement on solutions at spatial
infinity.  \\ 
In this framework, in fact, the symplectic structure of the field theory is uniquely defined by
the bulk term, generated by the so called \textit{reduced symplectic current}. There is
no contribution to the symplectic structure
from the boundary terms, which are instead pushed directly  into the very definition of the
variation of the Hamiltonian. \\
The variation of energy, as previously remarked, is defined as the on-shell variation of the
Hamiltonian. In this framework different definitions of energy for the same system do in fact
arise in correspondence to different definitions of the variations of the Hamiltonian  (and
consequently of the variations of energy). Different variations are related to the different
vertical vector fields which generate different one-parameter families of
solutions starting from the same solution. Different one-parameter families correspond to
different boundary conditions the solutions have to satisfy. The variational equation which
defines the variation of energy can be (or can be not) integrated, depending on
the boundary conditions imposed, when integrable different kinds of energy are defined. \\
This viewpoint is in full accordance with the classical treatment of thermodynamical
systems. When we impose
different boundary conditions on the same thermodynamical system, in fact we perform each time a
different choice between the intensive or extensive variables for the system (in a symplectic
framework we can say that we are choosing the "\textit{control parameters}") and correspondingly
we expect to find different energies for the system. \\ 
In  vacuum General Relativity, as  shown in \cite{raiteri}, the
definition of quasilocal energy via Noether theorem leads to the same results previously obtained
 by using the Trace-K action functional  \cite{BLY}, \cite{BY} or by using a Hamiltonian
analysis as in  \cite{Booth},  \cite{Kijowski}. 
The Hilbert Lagrangian is sensitive to the formalism presented here. In presence of 
non-orthogonal boundaries, in fact, the generalized Regge-Teitelboim correction term $\tau$ (we
add to the definition of $\delta H $) is fundamental for the definition of the variation of
energy of the gravitational field. In this way a definition of quasilocal internal energy for
spatially bounded gravitating systems can be obtained where the reference background is properly
taken into account. This is obtained by imposing Dirichlet boundary conditions, i.e.  the metric
at the boundary of the world tube is kept fixed.\\  In this paper we apply our general recipe
\cite{raiteri} to the case of Einstein-Maxwell theory. Electromagnetism is here treated as a
natural theory. In natural theories each vector field tangent to the spacetime manifold can be
naturally lifted to a vector field tangent to the
configuration bundle and moreover this lifted vector field is a symmetry
for the
Lagrangian \cite{Lagrange}. We remark that in the natural lift of spacetime vector fields no gauge
freedom remains undetermined (quite different is the case in gauge theories!). In this well
defined geometric framework the Lie derivative of the fields (i.e. the electromagnetic potential)
is uniquely defined, has a correct mathematical interpretation and it is thence possible to define
Noether conserved quantities via Noether theorem. \\ 
Maxwell theory,  opposite to other gauge theories, can be treated as a natural theory by using
a suitable representation of the gauge group $U(1)$. The configuration bundle turns out to be a
\textit{natural} $U(1)$  bundle; this in turn implies that the configuration bundle is a 
trivial $U(1)$  bundle, meaning that no magnetic charge is allowed: see \cite{Robutti}. \\
In this particular framework the electromagnetic potential is a geometric object of order $2$ and
it is possible   to apply the formula which defines the correct variation of energy
if we introduce also an electromagnetic correction term $\tau$ (related to the 
representation chosen) in the definition of the variation of the Hamiltonian. This
allows to define correctly the energy for the system. Moreover we obtain
the symplectic structure and the phase space of the theory, with results in full accordance
with \cite{Wald}. \\
To perform the integration of the variational equation which defines the variation of energy for
the electromagnetic field we choose two different sets of boundary conditions, corresponding
respectively, to an adiabatic system and to an electrically isolated system \cite{Kijowski}. In
both these cases we obtain a priori a suitable energy contribution to the Einstein-Maxwell
Lagragian due to the pure electromagnetic field. In the first case this contribution vanishes,
this meaning that the contribution to the energy comes out only from the pure gravitational part
of the Noether charge and it keeps track of the electromagnetic field only through the solutions
of Einstein-Maxwell field equations. In the second case the electromagnetic contribution to energy
turns out instead to be  proportional to the product of the electric charge and the electrostatic
potential (integrated over the boundary), a result which is
physically reliable if compared with Classical Electrodynamics \cite{elettrodinamica}. \\

Gluing together the results obtained for the gravitational field and for the
electromagnetic field, we can calculate the energy content of a spatially
bounded region of spacetime in the framework of Einstein-Maxwell theory and it is then possible to
apply the  theory developed so far to the case of rigidly rotating horizons  (see
\cite{Ashtekar2},
\cite{Booth}).\\ 
The classical definition of black hole thermodynamics, based on the
definition of entropy  for a Killing horizon, deals in fact with quite  unphysical models, as it
applies only to  static or quasi-static spacetimes (which means small perturbations from a static
situation and thus no radiation is admitted nearby the horizon). Moreover, to define the concept
of event horizon for non-stationary spacetimes, we need to know the whole history of the
spacetime and this is in contrast with the concept of physical observer.\\
A generalization of these concepts to more physical situations has then been proposed by
Ashtekar and his coworkers in \cite{Ashtekar1}, \cite{Ashtekar2} (and
references therein), where they introduce the notion of
isolated horizon as a $3$-dimensional null hypersurface $\Delta$ embedded into spacetime.
Cross sections of isolated horizons are, roughly speaking, non-expanding surfaces, isolated
from the outside and with a null flux of matter and radiation through or outside them. To define
the variation of energy and the conserved quantities of isolated horizons in the Noether framework
we do not need that the spacetime  has a
global Killing vector field but it is sufficient to assume the existence of a local Killing vector
field for the $3$-metric of the horizon;  this geometric requirement is fulfilled by
isolated horizons. Indeed the horizon Killing vector field ensures that isolated horizon
are (quasi) locally in equilibrium, but they are allowed to  admit nearby  radiation.\\
The Noether formalism developed to define quasilocal conserved
quantities for Einstein-Maxwell theory naturally applies to isolated horizons and in
particular we can define the area $A_\Delta$, the angular momentum $J_\Delta$ and the charge
$Q_\Delta$ of the horizon through integrals
on the cross sections $\Delta_t$ of $\Delta$ and they are conserved on the whole
horizon. From the definition of the variation of energy we can
obtain a first principle for rigidly rotating horizons, which are defined as (weakly) isolated
horizons with an internal symmetry, generated by a vector field tangent to the cross sections.
This requirement together with the definition of (weakly) isolated horizons, ensures that when we
evaluate the variation of energy on a cross section $\Delta_t$, a first law of black holes
thermodynamics is defined under the form:
\be
\delta E_{\Delta}=\frac{k_{(l)}}{k} \delta A_\Delta+ \Phi_{(l)} \delta Q_\Delta+
\Omega_{(l)} \delta J_\Delta 
\ee
where $k_{(l)}$, $\Phi_{(l)}$, $\Omega_{(l)}$ are parameters of the horizon related respectivly to
its temperature, its electrostatic potential and its angular velocity.\\

This paper is divided into six Sections. In Section $2$ we  review 
the definition of the Hamiltonian structure of a field theory and  the definition of the
variation of energy, via Noether theorem. The formalism used is a pure geometric
approach, based on the
definition of Lagrangian field theories on jet-bundles \cite{Katz}. In
Section $3$ we apply the formalism to the case of 
 General Relativity and we find an explicit expression for the variation of
energy of spatially bounded gravitating systems. We introduce here all
concepts and notation that are useful for a $(3+1)$ approach to field theories. In Section $4$
we apply our definitions to the electromagnetic field and we analyse the  definition of variation
of energy for Einstein-Maxwell theory. The general theory is applied  to
calculate the energy for two particular sets of boundary conditions,
corresponding to different physical situations. 
In Section $5$ we finally introduce the boundary and geometric conditions
which
define an isolated and a rigidly rotating horizon. The direct evaluation of the
formula defining the variation of energy on such surfaces is nothing but the first law
 of thermodynamics for rigidly rotating horizons.

\section{\large Geometric framework}

The definition of the Hamiltonian for a field theory, as well as the
definition of energy
as its on shell value, can be based on Noether's theorem. It has been
shown in \cite{raiteri} that the formula obtained via this definition reproduces, in
applications,  the same results obtained in the ADM  Hamiltonian formalism \cite{Booth},
\cite{Kijowski} or obtained using a Hamilton-Jacobi analysis of the Trace-K action functional 
 \cite{BLY}, \cite{BY}. The advantages arising from Noether formalism are related to the fact that
the Noether (covariantly) conserved quantities are independent on the addition of
boundary terms to the Lagrangian and consequently these latter terms do not influence the
definition of energy.
In the original analysis of Brown and York \cite{BY} different
boundary terms for
the
action functional are dictated by different boundary conditions and
 by different choices of the control modes of the boundary data. Before entering into the
detais of the matter, we shall shortly summarize the geometric framework we shall need in the
rest of the paper (we address the reader to \cite{Lagrange}, \cite{Katz}, \cite{Kolar} for a
deeper and more rigorous mathematical exposition).\\ 
The \textit{configuration bundle} for a Lagrangian field theory on a manifold $M$, with dimension
dim$(M)$=$m$, is a bundle $(B,M;\pi)$. A \textit{field configuration} is a
section $\sigma$ of $B$; i.e. a map $ \sigma: M \rightarrow B$. We can choose fibered coordinates
  $(x^\mu, y^{i}) $ on $B$ and, in these coordinates, the field configuration can be
locally  represented as $\sigma^{i}: x^\mu \mapsto (x^\mu, y^{i} =\sigma^{i} (x^\mu))$.
The jet prolongations of the configuration bundle are denoted by $(J^k
B,M;j^k \pi)$. Local fibered coordinates $(x^\mu, y^{i}
,y^{i}_{\mu},..,y^{i}_{\mu_1..\mu_k})$ can be chosen on $J^k B$ and they denote the spacetime
derivatives of fields up to order $k$. The prolongation of a field configuration $\sigma$ to a jet
bundle is denoted by $j^k \sigma$. \\
A \textit{vertical vector field} is a section $X$ of the vertical bundle
$V(B)= \text{Ker} (T \pi)$, namely a vectorfield which is everywhere tangent to the fibers;
locally it can be written in fiber coordinates as $X=X^{i} \frac{\pa}{\pa y^{i}} \equiv \delta
y^{i}
\frac{\pa}{\pa y^{i}}$ and it describes the variation  $\delta y^{i}$ of the dynamical fields. Its
prolongation to the $k$-order jet bundle of
$B$ is denoted by $j^k X$ and describes the variation of the fields and of their derivatives up to
order $k$. If we denote by $\Phi_t$ the $1$-parameter flow generated by $X$ on $B$, its
prolongation to the jet bundle $J^k B$ defines a flow on it, coherently denoted by $j^k \Phi_t$.\\
The variation of a generic morphism $R: J^k B \rightarrow \Lambda^n  M$ (where $\Lambda^n  M$
denotes the $n$-form bundle over $M$ for  $n \leq m$), along the flow of $X$ at any given section
$\sigma$ can be defined as:
\begin{eqnarray}
(\delta_X R)(\sigma)=\frac{d}{dt}
R \Big( j^k \Phi_t  \circ j^k \sigma  \Big)\Big|_{t=0}=
\frac{d}{dt}
R \Big( j^k  \sigma_t  \Big)\Big|_{t=0} \label{qze}
\end{eqnarray}
where $\sigma_t=\Phi_t  \circ \sigma$ denotes a $1$-parameter family of field
configurations on $B$ obtained by dragging $\sigma$ along the flow of $X$. \\
A $k$-order \textit{Lagrangian}  is a morphism from the
$k$-order jet bundle
$J^k B$ to the bundle $\Lambda^m (M)$ of volume $m$-forms on $M$ and locally it can be
written as
$L=\c{L} (y^{i} ,y^{i}_{\mu_1},..,y^{i}_{\mu_1..\mu_k}) ds $, where
$\mathfrak{L}$ is  scalar density called the \textit{Lagrangian scalar density}
and $ds= dx^1 \we d x^2 \we.. \we d x^m$ is the local volume $m$-form on
$M$ (we do not assume an explicit dependence of the Lagrangian on spacetime coordinates
$x^\lambda$ because this dependence is forbidden in natural theories, see \cite{Lagrange}).\\
 The \textit{action functional}
\begin{equation}
A_D(\sigma)=\int_D (j^k\sigma)^* L \label{action}
\end{equation}
is defined by integrating the pull-back $(j^k \sigma)^* L$ of the Lagrangian $L$ over a
compact region $D\subset M$ with regular boundary $\pa D$.
According to Hamilton's principle, field equations are obtained by imposing the
action $A_D (\sigma)$ to be stationary along the flow generated by any
compactly supported
vertical field $X \in V(B)$. In this way one obtains Euler-Lagrange field
equations $(\mathbb{E} \circ j^{2 k}\sigma)=0$ and sections $\sigma$  of $B$ which satisfy
this
variational
equation are called \textit{critical sections} (see \cite{Lagrange}, 
\cite{Katz} and \cite{Trautman}).\\

\noindent The variation of the Lagrangian along the $1$-parameter flow of the
vertical vector field $X$ can be defined as a global bundle morphism
$\delta L: J^k B \to V^* (J^k B)\otimes  \Lambda^m (M) $ by:
\begin{eqnarray}
<\delta L \circ j^k \sigma \mid j^k X>=\frac{d}{dt} \Big( L  \circ
 j^k \Phi_t  \circ j^k \sigma  \Big)\Big|_{t=0}
\end{eqnarray}
where $V^*( \cdot)$ denotes the dual bundle of the vertical bundle $V(\cdot)$ and $< \cdot \mid
 \cdot>$ denotes the natural duality between $V^*( \cdot)$ and $V( \cdot)$.\\ 
In this paper we consider only \textit{natural theories},
namely those field theories for which the configuration bundle $B$ is natural, i.e.\ each
spacetime vector field $\xi=\xi^\mu \pa_\mu$ lifts naturally (i.e. preserving the
commutators)  to a unique vector field $\hat\xi$ over the configuration
bundle as well as it uniquely lifts to any  prolongation $J^k B$.
This means that for each $\xi =\xi^\mu  \pa_\mu \in \mathfrak{X}(M)$
we have a naturally lifted vector
field $\hat{\xi} \in \mathfrak{X}(B)$ defined by:
\begin{eqnarray}
\hat{\xi} =\xi^\mu(x^\mu) \pa_\mu+\xi^i(x^\mu, y^{j}) \pa_i \label{vect}
\end{eqnarray}
while $j^k \hat{\xi}$ is naturally defined as the prolongation
of the vector field $\hat{\xi}$ to the jet bundle $J^k B$. 
We stress that the coefficients $\xi^{i}$ in (\ref{vect}) depend on $\xi^\mu$ and their
derivatives up to an algorithmically computable finite order $r$, which depends on the case
considered, called the
\textit{total order of the natural bundle}. The sum
$s=k+r$ of the order of the Lagrangian with the order of the natural bundle is called the
\textit{total order of the theory}.\\
A Lagrangian theory is \textit{natural} or, "physically" speaking,
\textit{covariant} if each spacetime vector field $\xi$ is an
\textit{infinitesimal Lagrangian symmetry}, which is equivalent to state that the following holds:
\begin{eqnarray}
\pounds_{\xi} L=<\delta L  \mid j^k \pounds_{{\xi}} \sigma> \label{simmetr}
\end{eqnarray}
where $\pounds_{{\xi}} \sigma$ is the \textit{Lie derivative of a
section $\sigma$} defined by
\be
\pounds_{{\xi}} \sigma=T \sigma (\xi)-\hat{\xi}  \circ
\sigma\equiv(\pounds_{{\xi}} y^{i})\pa_i
\ee
From this formula it is clear that $\pounds_{{\xi}} \sigma$ is a vertical
vector field over $B$ and it describes the evolution of the field configuration
$\sigma$ along the flow of $\xi$. In local coordinates formula
(\ref{simmetr}) is
equivalent to
require that:
\be
\pounds_{\xi} L= \Big( \frac{\pa L}{\pa y^{i}}    \pounds_{{\xi}} y^{i}+
\frac{\pa L}{\pa y^{i}_{\mu_1}}    \pounds_{{\xi}} y^{i}_{\mu_1}+..+
\frac{\pa L}{\pa y^{i}_{\mu_1..\mu_k} }    \pounds_{{\xi}}
y^{i}_{\mu_1..\mu_k}
\Big) ds
\ee
It is well known that General Relativity with the Hilbert Lagrangian is a covariant theory, i.e.
it is natural; we will show in the sequel that also electromagnetism with the
Einstein-Maxwell Lagrangian can be treated as a natural theory, provided the
configuration bundle and the Lie derivatives of the electromagnetic vector
potential $A_\mu$ are defined in an appropriate way   \cite{CADM}, \cite{Fnatu},
\cite{Kolar}.\\
The variation  $\delta L$  of the Lagrangian splits as follows  by a covariant
integration by parts:
\be
<\delta L \circ j^{k} \sigma\mid j^{k} X>=<\mathbb{E} (L) \circ j^{2 k} \sigma \mid
X> + \text{d} <\Bbb{F} (L) \circ j^{2k-1} \sigma \mid j^{k-1} X>
\label{dL}
\ee
where $\Bbb{E}$ and $\Bbb{F}$ are called respectively the
\textit{Euler-Lagrange}  and the \textit{Poincar\e-Cartan morphism} \cite{Lagrange}. We stress
that the  Poincar\e-Cartan morphism is not unique for theories of order higher than $2$  (even
though in natural theories uniqueness is achieved, for any order $k$, through the introduction of
a dynamical spacetime connection; see \cite{Remarks} and \cite{Sinicco} for
details).
When evaluated on a particular configuration and on a vertical vector field $X$, the morphisms 
$\Bbb{E}$ and $\Bbb{F}$ are respectively identified with a $m$-form and a $(m-1)$-form
over $M$. In local coordinates $\delta L$ can be expressed as:
\be
\delta L ( j^{ k} \sigma , j^{k} X)=\Bbb{E}_i ( j^{2 k} \sigma) X^{ i} ds+
\Bbb{F}^\lambda ( j^{2 k-1} \sigma, j^{k-1} X) ds_\lambda \label{eqmoto}
\ee
Substituting this expression into the variation $\delta_X A_D (\sigma)$ of the action functional
(\ref{action}), it is easy to see that $\Bbb{E}_i$ represent the Euler-Lagrange equations of
motion, while $\Bbb{F}^\lambda$ is a boundary term which vanishes if suitable
conditions are imposed on $X$ (usually one states that $X$ is vanishing on the
boundary $\pa D$ together with all its derivatives, which is equivalent to
state that it is "strongly"  compactly supported).\\
If we deal with natural field theories, Noether's theorem allows us to
construct a covariantly conserved current and a conserved quantity for each
vector field $\xi$ on $M$. This object can be globally constructed and well
defined
from a geometric point of view   \cite{Robutti},  \cite{Lagrange}. The
Noether current, for a solution $\sigma$ is locally defined as a $(m-1)$-form over $M$:
\be
\c{E}^\lambda  (L, \xi, \sigma ) =\Bbb{F}^\lambda ( j^{2 k-1} \sigma,
j^{k-1} (\pounds_{\xi}
\sigma))- \xi^\lambda \c{L} ( j^{ k} \sigma) \label{corr}
\ee
It is well known that there exists a unique and global decomposition of the
Noether current as the sum of a $(m-1)$-form $\tilde{\c{E}}$ of $M$ (called the
\textit{reduced current}, which vanishes on shell) and the divergence of a
$(m-2)$-form $\c{U}$, named the \textit{superpotential} of the  theory (see
 \cite{Lagrange}):
\be
\c{E} (L, \xi, \sigma ) =\c{E}^\lambda  (L, \xi, \sigma )
ds_\lambda=\tilde{\c{E}}^\lambda  (L, \xi,
\sigma ) ds_\lambda+ [ d_\mu {\c{U}}^{ \lambda \mu}  (L, \xi, \sigma)] ds_{
\lambda} \label{corcon}
\ee
Both the superpotential and the reduced current are linear in the components of the
infinitesimal simmetry generator $\xi$ and their derivatives up to order $(k+r-2)$ and $(k+r-1)$
respectively, where $k+r$ is the total order of the theory (see our discussion about formula
(\ref{vect})).\\
Covariantly conserved quantities are naturally defined as the integral of the current
$\c{E}$ on a
$(m-1)$-dimensional
submanifold $\Sigma \subset M$ with a compact boundary $\pa \Sigma \subset
\Sigma \subset M$:
\be
Q_\Sigma (L, \xi, \sigma ) =\int_ \Sigma {\c{E}}^\lambda  (L, \xi, \sigma
) ds_\lambda= \int_ \Sigma \tilde{\c{E}}^\lambda  (L, \xi, \sigma
) ds_\lambda+\frac{1}{2} \int_{\pa \Sigma}   {\c{U}}^{\mu \lambda}  (L, \xi,
\sigma)  ds_{ \mu \lambda}
\ee
When evaluated on shell, since $\tilde{\c{E}}$ is proportional to the equations of motion, the
conserved quantities are  pure boundary terms integrated on $\pa \Sigma$:
\be
Q_\Sigma (L, \xi, \sigma ) =\frac{1}{2} \int_{\pa \Sigma}   {\c{U}}^{\lambda \mu}  (L,
\xi,\sigma)  ds_{\lambda \mu}
\ee
which means (see (\ref{corcon})) that the conserved Noether currents are  exact on-shell. When
 evaluated for a particular theory and for a fixed
$\xi$ these quantities do not reproduce the expected physical values because of the
"anomalous factor problem" \cite{RT}. The solution to this problem relies on the so called
\textit{ADM covariant formalism}  \cite{CADM}, \cite{Sinicco}. According to
this, we compute the variation of the conserved current (\ref{corr}) along a vertical vector field
$X$ as:
\ba
\delta_X {\c{E}}  (L, \xi, \sigma)&=& \delta_X \Bbb{F} ( L, \pounds_{\xi}
\sigma)-i_\xi (\delta_X  L) =\nonumber\\
&=&\delta_X \Bbb{F} (  L, \pounds_{\xi}
\sigma)-i_\xi \Bbb{E} (L,X)- i_\xi [d\Bbb{F} ( L, \pounds_{\xi}
\sigma) ] \nonumber=\\
&=&\delta_X \Bbb{F} (  L,\pounds_{\xi}
\sigma)-\pounds_\xi \Bbb{F} (  L,X)-i_\xi \Bbb{E} (L,X)+d [  i_\xi \Bbb{F} ( L,\pounds_{\xi}
\sigma) ] \nonumber=\\
&=&\omega ( L, X, \pounds_{\xi}
\sigma)- i_\xi \Bbb{E} (L,X)+ \text{d} (i_\xi \Bbb{F} (L,X))\label{trev}
\ea
where $\delta_X$ is defined by formula (\ref{qze}) and we shortly set $\Bbb{F} ( L, \pounds_{\xi}
\sigma)=\Bbb{F}^\lambda ( j^{2 k-1} \sigma, j^{k-1} (\pounds_{\xi} \sigma)) d s_\lambda$. 
The \textit{symplectic current} $\omega$  (see \cite{Wald}) is a $(m-1)$-form over $M$
\be
\omega ( L, X, \pounds_{\xi}\sigma)= \delta_X \Bbb{F} (L, \pounds_{\xi}
\sigma)-  \pounds_{\xi} \Bbb{F} (L, X)\label{cusy}
\ee
which suitably defines a symplectic structure for the field theory.
The covariant ADM method consists in defining the variation of the conserved
quantity, pushing the boundary terms appearing in the right hand side of (\ref{trev}) into the
definition itself, namely:
\ba
\delta_X \hat{Q}_\Sigma (L, \xi, \sigma ) &=&
\int_{ \Sigma}  \delta_X
{\c{E}}  (L, \xi, \sigma )- \int_{\pa \Sigma} i_\xi \Bbb{F} (L,X)= \label{qcorr}\\
&=&\int_{ \Sigma}  \delta_X
\tilde{\c{E}}  (L, \xi, \sigma )+ \int_{\pa \Sigma}  [ \delta_X
\c{U} (L, \xi )-i_\xi \Bbb{F} (L,X)] = \nonumber\\
&=&\int_{ \Sigma} \omega ( L, X, (\pounds_{\xi}\sigma))-\int_{ \Sigma} i_\xi \Bbb{E}
(L,X)\nonumber
\ea
This definition generalises to all natural theories the analysis originally given by Regge and
Teitelboim \cite{RT} for the ADM Hamiltonian with asymptotically flat solutions and gives 
the physically expected results for all the conserved quantities of the
theory.\\

The Hamiltonian structure of the theory naturally arises from this definition
\cite{Booth}, \cite{CADM}, \cite{raiteri}, \cite{Kijowski}. We
define the \textit{variation of the Hamiltonian} as the variation of the
conserved quantity, relative to a Cauchy surface $\Sigma$. The fundamental
requirement is that the infinitesimal generator $\xi$ of the symmetry,
which
defines $\delta H$, is transverse to the surface
$\Sigma$. From equation (\ref{qcorr}) we have that:
\ba
\delta_X \hat{H} (L, \xi, \sigma ) &\equiv &
\int_{ \Sigma}  \delta_X
{\c{E}}  (L, \xi, \sigma )- \int_{\pa \Sigma} i_\xi \Bbb{F} (L,X)= \nonumber\\
&=&\int_{ \Sigma}   \omega ( L, X,
\pounds_{\xi}\sigma)-\int_{ \Sigma} i_\xi \Bbb{E} (L,X) \label{htilde}
\ea
The variation of the energy is defined as the on-shell
variation of the Hamiltonian  \cite{Booth}, \cite{Kijowski}; in this case
$\Bbb{E} (L,X)$ vanishes and the variation of energy  is defined as the integral of the
symplectic current $\omega$ on $\Sigma$. We remark that if both $X$ and $\pounds_\xi \sigma$ are
solutions of the linearized field equations, from (\ref{cusy}) we have that $\omega$ is closed
on-shell \cite{Wald}, \cite{Remarks}.\\ 
The Poincar\e-Cartan form is linear in $X$ and its derivatives up to order $(k-1)$ and is linear
in  $\xi$ and its derivatives up to order $(k+r-1)$, where $r$ is the order of the natural
bundle. When we integrate (\ref{cusy})  on $\Sigma$ it is then possible to split the symplectic
current into two terms (through an integration by parts):
\be
\int_{ \Sigma} \omega ( L, X,\pounds_{\xi}\sigma)=\int_{ \Sigma}  \tilde{\omega} ( L,
X,\pounds_{\xi}\sigma)+\int_{ \Sigma} \text{d}  [\tau ( L, X,\pounds_{\xi}\sigma)]
\label{symp}
\ee
where $\tilde{\omega}$ is a $(m-1)$-form, while $\tau $ is a $(m-2)$-form which can be integrated
on the boundary $\pa \Sigma$ using Stokes' theorem\footnote{
It  is the analogy  with Classical Mechanics  which
suggests  how to perform the splitting:  roughly speaking, 
our purpose is to generalize  to natural field theories  the
well-known formula
$\delta H= \dot q\,\delta p-\dot p\,\delta q$ for Hamilton equations
in Classical Mechanics. Hence, in the integral $\int_\Sigma  
\omega (L,X,\pounds_\xi
\sigma)$ all the terms  which are not of the form $\int_\Sigma
\left[(\pounds_\xi \sigma)\, \delta_X p-(\pounds_\xi
p)\, \delta_X \sigma \right]d^3x$  have to be pushed, through integration by
parts, into the boundary term $\int_{\partial \Sigma}
\tau(L,X,\pounds_\xi \sigma)$.}. Our aim is thence to suitably correct the
definition of the Hamiltonian and consequently of energy in
a way that the symplectic structure of the theory is completely defined by $\int_\Sigma
\tilde{\omega}$, i.e. by an integral over the $(m-1)$-surface $\Sigma$, while energy is a pure
boundary term, i.e. an integral over $\pa \Sigma$.\\ 
According with the prescription of the covariant ADM method
we can redefine a corrected variation of the Hamiltonian
$\delta_X {H}$ by pushing the new boundary term $\tau$ into the definition of $\delta_X H$ itself,
i.e.:
\ba
\delta_X {H} (L, \xi, \Sigma )&=&
\delta_X \hat{H} (L, \xi, \Sigma )-\int_{\pa \Sigma} \tau (
L,X,\pounds_{\xi}\sigma)=\\
&=&\int_{ \Sigma}  \delta_X
\tilde{\c{E}}  (L, \xi, \sigma )+ \int_{\pa \Sigma} [ \delta_X
\c{U} (L, \xi )-i_\xi \Bbb{F} (L,X)-\tau ( L,
X,\pounds_{\xi}\sigma) ] \label{varacca}
\ea
The new variation of the Hamiltonian does no longer contain any boundary term. In fact
boundary terms arising from the variation of the reduced current $ \delta_X
\tilde{\c{E}}$ are completely
cancelled by the variation of the superpotential together with the
correction terms $i_\xi \Bbb{F}$ and $\tau$, so that, from equations (\ref{htilde}) and
(\ref{symp}),
$\delta_X H$ finally results to be:
\be
\delta_X {H}(L, \xi, \Sigma )=
\int_{ \Sigma}  \tilde{\omega} ( L, X,
\pounds_{\xi}\sigma)-\int_{ \Sigma} i_\xi \Bbb{E} (L,X) \label{etilde}
\ee
We can now define the \textit{ variation of energy} as the on-shell value of the
variation
of the Hamiltonian \cite{raiteri}. It turns out to be:
\be
\delta_X E (L, \xi) = \int_{\pa \Sigma} [\delta_X \c{U} (L, \xi )-i_\xi \Bbb{F}
(L,X)-\tau ( L, X,\pounds_{\xi}\sigma)] \label{ENE}
\ee
because $\delta_X \tilde{\c{E}}$ in (\ref{varacca}) is identically zero since we have assumed that
$X$ be a solution of the linearized equations of motion. This \textit{master formula}
(\ref{ENE}) gives us a recipe to define the energy $E(L, \xi)$ once the variational equation
$\delta_X E$ is solved. We stress that the definition (\ref{varacca}) of the variation of the
Hamiltonian  and consequently the definition (\ref{ENE}) of energy variation
$\delta E$ do not depend on any divergence term possibly added to the Lagrangian.\\
We indeed remark that the final formula of $\delta H$ is invariant under a change
of the Lagrangian by pure divergence terms. This means that the variation of the Hamiltonian
and the variation of energy are defined for an equivalence class $[L]$ of Lagrangians,
rather than for a single one, defined by the equivalence relation:
\baa
L\sim L' \Leftrightarrow \exists {T} \in \Lambda^{m-1} (M) \Rightarrow L=L'+ \text{d} {T}
\eaa
which implies that $L$ and $L'$ generate the same field equations. Alternative definitions of
energy that can be found in literature define $\delta H$ by adding some
boundary terms  to the Lagrangian or to the canonical Hamiltonian \cite{Booth}, \cite{BY},
\cite{HHnon}. These boundary terms which do not affect the equations of motion modify the
definition of energy and consequently the control mode of the fields on the boundary
\cite{Kijowski}. Suitable terms have then to be added a posteriori to
the action functional to obtain the predefined boundary control mode. \\
The symplectic structure of the theory and consequently its phase space are
instead defined in literature starting from the symplectic current $\tilde{\omega}$ defined on the
surface $\Sigma$ and by a symplectic structure on the boundary $\pa \Sigma$, which is
dictated by the choice of the boundary term we add to the Lagrangian; see \cite{Ashtekar2}.\\
The definition of the variation of the Hamiltonian we
propose gets rid of all these problems, since the symplectic structure is
defined only by the
surface part $\int_\Sigma \tilde{\omega}$ of the symplectic current. In analogy with
Classical Mechanics, as shown in \cite{Ashtekar2}, \cite{raiteri}, \cite{Kijowski},
the variation of the Hamiltonian contains a term which determines the symplectic structure of the
theory as well as a term which contains the Lagrangian equations of
motion. This is just the case of (\ref{etilde}), where $\tilde{\omega}$ is related to the
symplectic structure while the second term is linear in the Lagrangian equations of
motion so that, eventually, formula (\ref{varacca}) can be considered as a well-defined
expression for the variation of the Hamiltonian.\\ 
Moreover, boundary conditions in this framework do not
play any role either in the definition
of the
variation of energy nor in the definition of the symplectic structure of the
theory. They will instead play a key role in the integration of the
variational equation $\delta_X E$ where imposing boundary conditions means to fix the values
of the vertical vector field $X$ (i.e. of the variations of fields) at the boundary
$\pa \Sigma$. The choice of boundary conditions, namely the choice of the vertical vector field
$X$, together with a reference "zero level" for energy, provides us with the definition of
energy  
\cite{BLY}, \cite{raiteri}.\\
We will see that this formalism applies to Einstein gravitational
field and also to the electromagnetic field in the framework of a geometric natural formulation
of  Einstein-Maxwell theory. This is possible as the total order of the electromagnetic theory
results to be $s=3$.
\\ The theory has been formulated to calculate energy for a bounded,
compact region of spacetime. However it is possible to consider the case of a  system of infinite
spatial extent, i.e. $\pa \Sigma$ in (\ref{ENE}) is now the spatial infinity. Results obtained
using the quasilocal definition of conserved quantities reproduce the
results obtained through the standard ADM definitions. Moreover the
definition of the variation of energy (\ref{ENE}) is homologically invariant, if $\xi$ is a
global Killing vector field \cite{raiteri}. In the case that spatial infinity is homologic to a
finite bounded surface $B$, conserved quantities evaluated on $B$ correspond to the total
conserved quantities at spatial infinity.


\section{ General Relativity}
The main application of the formalism exposed in the previous chapter is General
Relativity in vacuum with dim$M$=4. The Lagrangian of the theory is the Hilbert Lagrangian:
\be
L_H=\frac{1}{2k } \sqrt{g} g^{\mu \nu}   R_{\mu \nu} ds
\ee
where $k=8 \pi $ in geometric units, with $G=c=1$. This argument has been analysed in a previous
paper \cite{raiteri} (for details on calculations and notations see also \cite{BLY}); we
give here just
a brief overview to recall the main results. 
We consider a four dimensional manifold $D\subset M $ of spacetime and we
assume $D$ to
be diffeomorphic to the product
$\Sigma
\times \R$, here $\Sigma$ is a 3-dimensional closed manifold with boundary
$\pa \Sigma=B$. For any $t
\in \R$ there exists an immersion:
\ba
\phi_t: \Sigma \rightarrow \Sigma_t
\ea
and we require $\Sigma_t$ to be a portion of a spacelike Cauchy hypersurface. The set of all
 leaves $\Sigma_t$ defines a \textit{foliation} of $D$ in spacelike hypersurfaces.
 Each $\Sigma_t$ intersects $\pa D$ in a two dimensional surface $B_t$ which is
diffeomorphic to $B$. We denote by $\c{B}$ the timelike hypersurface $\c{B}= \pa D=\bigcup_t
B_t$.\\ 
The time evolution is  generated by a vector field $\xi$ transverse to
$\Sigma_t$; we can require that $\xi^\mu \n_\mu t=1$ and $\xi$ is tangent to the
boundary $\c{B}$. We denote by $u^\mu$ the timelike
future directed normal to
$\Sigma_t$ and by $n^\mu$ the outward pointing  spacelike  normal to  $B_t$ in
$\Sigma_t$. 
The vector field $\xi$ can be decomposed as
\ba
\xi^\mu= N u^\mu+N^\mu \label{tempo}
\ea
where the \textit{shift} $N^\mu$ lies in $T \Sigma_t$, i.e. $N^\mu u_\mu=0$ (this
matter will be discussed later in detail).\\
We define  $\bar{n}^\mu$ the outward pointing spacelike unit normal to
$\c{B}$ while by $\bar{u}^\mu$ we define the timelike future directed normal to
$B_t$ in $\c{B}$. Barred and unbarred vectors are related by boost relations (see \cite{BLY}):
\baa
\bar{u}^\mu=\gamma  {u}^\mu+\gamma  v {n}^\mu \\
\bar{n}^\mu=\gamma  {n}^\mu+\gamma  v {u}^\mu
\eaa
where we have set:
\ba
\gamma  v &=&\bar{u}_\mu {n}^\mu=-\bar{n}_\mu {u}^\mu=sinh (\theta )\nonumber \\
v&=&\frac{N^\mu n_\mu}{N}=-\frac{\xi^\mu n_\mu}{\xi^\mu u_\mu} \label{velboost}
\ea
The parameter $\theta$ is usually called the \textit{velocity parameter}, while
$v$ is the boost velocity and $\gamma=(1-v^2)^{-1/2}$.
In fact, if we consider a point on
$B_t$ at a time $t$ we can observe its evolution according to the unboosted (unbarred)
observers, which means observers at rest on $\Sigma_t$  and evolving with the
normal vector $u^\mu$. Otherwise we can describe its evolution with
respect to a boosted (barred) observer evolving with the four velocity $\xi^\mu$ and seeing
the vectors $\bar{u}^\mu, \bar{n}^\mu$ as normals to $B_t$. The scalar $v$ is
in relation to the
boost radial velocity between the two classes of observers, while $\theta$ describes the
non-orthogonality of the foliation; see \cite{BLY}. \\ 
The metrics induced on the surface $\Sigma_t$ and $\c{B}$ by the projector operators are defined
respectively as:
\baa
h_{\mu \nu}= g_{\mu \nu}+{u}_\mu {u}_\nu\\
\bar{\gamma}_{\mu \nu}= g_{\mu \nu}- \bar{n}_\mu \bar{n}_\nu
\eaa
while the metric on $B_t$ can be defined with respect to boosted or unboosted
observers as:
\ba
\sigma_{\mu \nu}= g_{\mu \nu}- \bar{n}_\mu \bar{n}_\nu+\bar{u}_\mu \bar{u}_\nu=
g_{\mu \nu}- {n}_\mu {n}_\nu+{u}_\mu {u}_\nu
\ea
In the sequel we denote by $g, h, \bar{\gamma}, \sigma$ the absolute values
of the corresponding metric determinants. The \textit{extrinsic curvatures} $K_{\mu \nu}$ of
$\Sigma_t$ in $M$,
$\bar{\Theta}_{\mu \nu}$ of $\c{B}$ in $M$ and $\c{K}_{\mu \nu}$ of
$B_t$ in $\Sigma_t$ are defined respectivly as:
\baa
K_{\mu \nu}= -{h}_\mu^\alpha \n_\alpha u_\nu \\
\bar{\Theta}_{\mu \nu}= -{\bar{\gamma}_\mu}^\alpha \n_\alpha \bar{n}_\nu\\
\c{K}_{\mu \nu}= -{\sigma}_\mu^\alpha D_\alpha n_\nu
\eaa
where $D$ is the metric covariant derivative with respect to the
metric $h$ of $\Sigma_t$ .
We denote by $K$, $\bar{\Theta}$, $\c{K}$ the above quantities contracted with
the corresponding metrics. We shall instead denote by  $\bar{\c{K}}_{\mu \nu}$ the extrinsic
curvature of $B_t$ with respect to the normal $\bar{n}^\mu$, i.e. $\bar{\c{K}}_{\mu \nu}=\gamma
\c{K}_{\mu \nu}-\gamma v {\sigma}_\mu^\alpha {\sigma}_\nu^\beta \n_\alpha u_\beta$; see
\cite{raiteri}.\\
The \textit{three momenta} $P^{\mu \nu}$ of the surfaces $\Sigma_t$ and $\Pi^{{\mu}
\nu}$ of $\c{B}$ are defined in terms of the extrinsic curvatures of the $3$-hypersurfaces
in $M$:
$$
\cases
{
P^{\mu \nu}=\frac{\sqrt{h}}{2 k} (K h^{\mu \nu}-K^{\mu \nu}) \cr
\bar{\Pi}^{{\mu} \nu}=-\frac{\sqrt{\bar{\gamma}}}{2 k} (\bar{\Theta} \bar{\gamma}^{\mu
\nu}-\bar{\Theta}^{\mu
\nu}) }
$$
The \textit{time evolution vector field} can be decomposed, into a normal
part $N$ to $\Sigma_t$ called \textit{lapse} and a tangent part $N^\mu$  called
\textit{shift}, by:
\baa
\xi^\mu =N u^\mu+N^\mu
\eaa
Equivalently it can be decomposed with respect to the boosted observers:
\baa
\xi^\mu =\bar{N} \bar{u}^\mu+ \bar{N}^\mu
\eaa
where $\bar{N}=\frac{N}{\gamma}$ and $\bar{N}^\mu$ is the projection of
${N}^\mu$ on $B_t$, i.e. $\bar{N}^\mu={\sigma}^\mu_\alpha {N}^\alpha$.\\


\subsection{\large Variation of the Hamiltonian in General Relativity}

It is now possible to calculate explicitly the symplectic current and the variation of the
Hamiltonian for General Relativity in vacuum, which is
governed by the Hilbert Lagrangian (details and calculations can be found in \cite{raiteri}).\\
On a leaf $\Sigma_t$ of the foliation, the reduced symplectic current $\tilde{\omega}$ and the
boundary part $\tau$ in (\ref{symp}), turns out to be:
\ba
\tilde{\omega} ( L_H,X,\pounds_{\xi} g)&= &[(\pounds_{\xi} h_{\mu \nu}) \delta_X
P^{\mu \nu}-(\pounds_{\xi} P^{\mu \nu} )\delta_X h_{\mu \nu}] d^3 x \label{tildomeg} \\
\tau ( L_H, X,\pounds_{\xi}  g)&=& \frac{1}{2k}[\pounds_{\xi} (\sqrt{\sigma} n_\mu
\delta_X u^\mu)- \delta_X (\sqrt{\sigma} n_\mu
\pounds_{\xi} u^\mu)]d^2 x
\ea
From these expressions for the reduced symplectic current $\tilde{\omega}$ and
$\tau$ it is clear that the symplectic structure and the phase space of the
theory are both completely determined on the Cauchy surface
$\Sigma_t$ in terms of the $3$-metric $h_{\mu \nu}$ and its conjugated momentum $P^{\mu \nu}$. \\
The variation of the Hamiltonian (\ref{varacca}) has been calculated in \cite{raiteri} and turns
out to be:
\be
\delta_X H (L_H,\xi,\Sigma_t)= \int_{\Sigma_t} \{
\c{H} \delta_X N+ \c{H}_\alpha \delta_X N^\alpha+[ h_{\mu \nu}]_\xi
\delta_X P^{\mu \nu}-[ P^{\mu \nu}]_\xi \delta_X h_{\mu \nu} \} d^3 x
\label{ddih}\\
\ee
where $[h_{\mu \nu}]_\xi$ and $[P^{\mu \nu}]_\xi$ follow directly from the
$3$-dimensional Einstein equations. Notice that no boundary term appears in (\ref{ddih}). 
Comparing (\ref{tildomeg}) with (\ref{ddih}) the Hamiltonian equations (\ref{etilde}) read as
follows:
$$
\cases
{
-\frac{\sqrt{h}}{ k}  G^{\mu \nu} u_\mu u_\nu=\c{H}=0\cr
-\frac{\sqrt{h}}{ k}  G^{\mu \nu} h_{\alpha \mu} u_\nu=\c{H}_\alpha=0\cr
\pounds_\xi h_{\mu \nu}=[h_{\mu \nu}]_\xi\cr
\frac{\sqrt{h}}{2 k}  G^{\alpha \beta} h_{\alpha}^\mu h_{\beta}^\nu=\pounds_\xi
P^{\mu
\nu}-[P^{\mu \nu}]_\xi=0
}
$$
For our later purposes it is not necessary to formally integrate (\ref{ddih}) and
 to give the explicit
expression of $H$ (see  \cite{BLY} and \cite{raiteri}  for details). Using the formalism
developed, in fact, one  can directly turn to calculate the time rate of change of the
Hamiltonian. It is easy to show that
the Hamiltonian is conserved along the flow of a vector field  $\xi$ iff
$\xi$ is a Killing vector  for the boundary metric on $\c{B}$, i.e. $\pounds_\xi
\bar{\gamma}_{\mu \nu}=0$.\\ 
The variation of the energy turns out to be equal to (see again \cite{raiteri}):
\ba
\delta_X E (L_H,\xi)&=& \int_{B_t} \Big{\{}  \bar{N} \delta_X ( \sqrt{\sigma}
\bar{\epsilon})
-\bar{N}^\alpha \delta_X (\sqrt{\sigma} \bar{j_\alpha})+\frac{\bar{N}
\sqrt{\sigma}}{ 2}
\bar{s}^{\mu \nu} \delta_X \sigma_{\mu \nu}\Big{\}} d^2 x+ \label{degr} \\
&+& \frac{1}{k}\int_{B_t} [\pounds_\xi
(\sqrt{\sigma}) \delta_X (\theta)- \delta_X (\sqrt{\sigma})
\pounds_\xi(\theta)]  d^2 x \nonumber
\ea
where we have set:
$$
\cases{
\bar{\epsilon}=\big( \frac{ 1 }{k} \big)  \bar{\c{K}}\cr
\bar{j}_\alpha=-\frac{2}{\sqrt{\bar{\gamma}} } \sigma_{\alpha \mu }
\bar{\Pi}^{\mu \nu} \bar{u}_{\nu} \cr
\bar{s}^{\mu \nu}=\frac{1}{k} [(\bar{n}^{\alpha} \bar{a}_{\alpha}) \sigma^{\mu
\nu}-\bar{\c{K}} \sigma^{\mu \nu}+\bar{\c{K}}^{\mu \nu}]\cr
\bar{a}^{\nu} =\bar{u}^{\mu} (\n_\mu \bar{u}^{\nu} ) }
$$

To integrate the  expression (\ref{degr}) we need to impose suitable boundary
conditions. C.-M. Chen and J. M. Nester have shown that there exist only
two different Hamiltonian boundary terms which correspond to covariant boundary conditions and
they are respectively the Dirichlet and the Neumann control, which define two different energies 
\cite{Nester}.
 The energy corresponding to Dirichlet   boundary conditions is usually accepted as the internal
energy; Dirichlet boundary conditions fix the $3$-metric
$\bar{\gamma}$ on the boundary $\c{B}$:
\be
\delta_X \bar{N} \mid_{\c{B}}=\delta_X \bar{N}^\mu \mid_{\c{B}}=\delta_X \sigma_{\mu \nu}
\mid_{\c{B}}=0
\label{Diri}
\ee
which can be considered as a restriction on the vector field $X$ and its flow.
This last conditions allow to integrate (\ref{degr}) and we obtain:
\be
E (L_H, g, B_t)-E_0 (L_H, g_0, B_t)= \int_{B_t} \Big{\{}   \sqrt{\sigma} (\bar{N}
\bar{\epsilon}-{\bar{N}}^\alpha
{\bar{j}}_\alpha)+
\frac{1}{k} \pounds_\xi (\sqrt{\sigma})  (\theta) \Big{\}}d^2 x \label{BLYr}
\ee
We stress that we are integrating along a $1$-parameter family of solutions all 
satisfying the same fixed boundary conditions (\ref{Diri}). The term $E_0 (g_0, B_t)$
corresponds to the quasilocal energy of a background solution $g_0$ inside the $1$-parameter
family and it becomes the "zero level" for energy. Owing to the properties $ \bar{N}=
\bar{N}_0
\mid_{\c{B}}$, $
\bar{N}^\alpha= \bar{N}_0^\alpha \mid_{\c{B}}$ and $\sigma_{\mu \nu} ={\sigma_0}_{\mu
\nu} \mid_{\c{B}}$, we obtain the equivalent
expression:
\be
E (L_H, g, B_t)= \int_{B_t} \Big{\{}   \sqrt{\sigma} [\bar{N}
(\bar{\epsilon}-\bar{\epsilon}_0)-{\bar{N}}^\alpha
({\bar{j}}_\alpha-{\bar{j_0}}_\alpha)]+
\frac{1}{k} \pounds_\xi (\sqrt{\sigma})  (\theta-\theta_0) \Big{\}} d^2 x \label{E-EO}
\ee
If we assume the vector field $\xi$ to be a Killing vector on the
boundary, this guarantees the conservation of energy in time and the term $\pounds_\xi
(\sqrt{\sigma}) $ identically vanishes, so that formula (\ref{E-EO}) agrees with the definition
of energy given in \cite{BLY}. 


\section{ Electromagnetic field}

In this Section we shall describe the Einstein-Maxwell theory and we shall calculate 
its energy in relation with different boundary conditions. The energy of
the system is related to the control-mode  we choose. It is possible to
imagine a "gedanken experiment" which reproduces in a laboratory the same
conditions described; controlled quantities and energy
have in this case an interpretation as physical osservables \cite{Booth}, \cite{Kijowski}. \\
The approach we use here is different from the the ones usually found in
literature. We treat electromagnetism, described by the Maxwell
Lagrangian, as a natural theory (\cite{CADM} and \cite{Fnatu}),  which
implies that we consider it as a $U(1)$-gauge theory based  on a natural principal bundle. This
choice implies that is possible to construct  a fiber bundle of geometric objects on spacetime
(describing the configurations of the electromagnetic field) which is a
principal bundle over $M$.
An immediate consequence of this fact is that the magnetic charge of the
configurations considered is identically zero. This is the same case
usually treated in literature to describe the Hamiltonian formalism for
Einstein-Maxwell theory using a trace-K action \cite{Booth}, \cite{BoothIH},
but in this latter case the vanishing of magnetic charge is assumed (somewhat equivalently) as a
restriction on the class of solutions.
\\

\subsection{Einstein-Maxwell theory}

We consider a Maxwell theory described as a gauge theory with gauge group
$U(1)$,  based on a fibred principal \textit{natural} bundle $(P, M, U(1); \pi)$, whose existence
 has been proven in \cite{Fnat} and \cite{Fnatu}. A configuration is a
section of the connection bundle
$\frac{J^1 (P)}{U(1)}$, which we denote by $A:M \rightarrow \frac{J^1 (P)}{U(1)}$ where  
$A=A_\mu (x) d x^\mu$; it is called
the \textit{quadripotential} of the theory. The coefficients $A_\mu$ are $u(1)$-valued,
where $u(1)= i \R$ is the Lie algebra of the Lie group. This means that $A_\mu$ can be
interpreted as a
$U(1)$-connection and its curvature $F_{\mu \nu}$ is  the field strength. We are particulary
interested in  the interaction between electromagnetic fields and gravity, with configuration
bundle $(\frac{J^1 (P)}{U(1)} \times $ Lor$(M),M;\pi)$ and described by the Einstein-Maxwell
Lagrangian density:
\be
\c{L}_{EM} (j^2 g, j^1 A)=\c{L}_{H} (j^2 g)+\c{L}_{M} (g, j^1 A)
\ee
with:
\be
\c{L}_{M} (g, j^1 A)=-\frac{1}{ 2k} \sqrt{g} g^{\alpha \mu} g^{\beta \nu}
F_{\alpha \beta}F_{\mu \nu} \label{Lemm}
\ee
We have  defined $F_{\alpha \beta}$ as the \textit{curvature} of the
quadripotential

\be
F=d A=\frac{1}{2} F_{\alpha \beta} d x^\alpha \we d x^\beta=\frac{1}{2} (\pa_\alpha
A_\beta-\pa_\beta A_\alpha)d x^\alpha \we d x^\beta
\ee
while the \textit{naive momentum} with respect to the curvature $F$ is
denoted by:
\be
f^{\alpha \beta} =-k \frac{\pa L_M}{ \pa F_{\alpha \beta}}=
\sqrt{g} g^{\alpha \mu} g^{\beta \nu} F_{\mu \nu}
\ee
The components of the curvature $F$ can be decomposed on the Cauchy surface $\Sigma_t$, hence
defining the electric flux density and the magnetic induction of the electromagnetic field: they
represent the physically observable quantities. The Lie derivative of the electromagnetic 
potential
$A_\mu$, according to \cite{CADM} and  \cite{Fnatu}, is defined as:

\be
\pounds_\xi A_\mu= \xi^\nu (\pa_\nu A_\mu)+A_\nu (\pa_\mu \xi^\nu)+q d_\mu
(d_\nu \xi^\nu)= \dot{A}_\mu+q d_\mu
(d_\nu \xi^\nu) \label{LieA}
\ee
where $\dot{A}_\mu$ is a shortcut for $\dot{A}_\mu=\xi^\nu (\pa_\nu A_\mu)+A_\nu (\pa_\mu
\xi^\nu)$ and  $q$ is a dimensional constant which represents the geometric charge of the
field $A_\mu$ carried by the  natural representation
\ba
R_q: X \in GL(m, \R) \rightarrow \text{exp} \Big( i \frac{q}{ e} \mid  det (X) \mid \Big) \in U(1)
\label{reppr}
\ea
(here $e<0$ is the unit charge of the electron). This representation (\ref{reppr}) defines the
principal bundle $(P, M, U(1);\pi)$ as a bundle associated to the frame bundle (or in other words
it is the representation which defines  $(P, M, U(1))$ as a natural bundle; see \cite{Fnat},
\cite{Fnatu}). Notice from (\ref{LieA}) that the bundle of connections
$\frac{J^1 (P)}{U(1)}$ turns out to be a natural bundle of order $r=2$. Indeed, the Lie
derivatives $\pounds_\xi A$ of its sections depend on the coefficients $\xi^\mu$ and their
derivatives up to order two. Being the Maxwell Lagrangian (\ref{Lemm}) of order $k=1$ the total
order of the electromagnetic theory turns out to be $s=3$.\\ 
The Maxwell Lagrangian will be considered separately for the sake of convenience; the
generalization
to the Einstein-Maxwell theory can be easily obtained recalling the results
obtained in Section $3$. Euler-Lagrange equations of motion follow for
$\c{L}_M$ immediately from the variational principle (\ref{eqmoto}): we obtain that
the electromagnetic field must satisfy the equation $J^\mu=0$, where
\be
J^\mu=-\frac{2}{ k} (\pa_\alpha f^{\mu \alpha})=-\frac{2}{ k} (\n_\alpha f^{\mu \alpha})
\ee
is called the \textit{current density}. We remark that the Poincar\e-Cartan form is obtained by the prescription
(\ref{dL}) as:
\be
\Bbb{F}^\lambda  (L_M, X)=\frac{2}{ k} f^{\mu \lambda} \delta_X A_\mu \label{FemM}
\ee
Using the definition of the Noether
current, from formula (\ref{corr}) it follows that:
\be
\c{E}^\lambda  (L_M, \xi )= \frac{2}{ k} f^{\lambda \beta} (\pounds_\xi
A_\beta)-
\xi^\lambda \c{L}
\ee
and from (\ref{LieA}) we obtain the natural splitting of the current into a reduced current term
and the divergence of the superpotential
\be
\c{E}^\lambda  (L_M, \xi )=\tilde{\c{E}}^\lambda (L_M, \xi )+ d_\mu
{\c{U}}^{ \lambda \mu}  (L_M, \xi)\label{Ece}
\ee
where in local coordinates we have
$$
\cases
{
\tilde{\c{E}}^\lambda  (L_M, \xi )=\c{T}^\lambda_\nu \xi^\nu - J^\lambda
A_\alpha \xi^\alpha- q J^{(\lambda} \delta^{\nu)}_\alpha \pa_\nu \xi^\alpha + q
\pa_\nu \big{\{} J^{[\lambda}
\delta^{\nu]}_\alpha \big{\}} \xi^\alpha\cr 
{\c{U}}^{\lambda \mu }  (L_M, \xi)=\frac{2}{ k}
[f^{\mu \lambda} ( q d_\alpha \xi^\alpha+ A_\alpha \xi^\alpha)]-q J^{[\lambda}
\delta^{\mu]}_\alpha  \xi^\alpha
 }
$$
having  defined  the \textit{stress-energy tensor} of the
electromagnetic field by setting:
\be
\c{T}^\lambda_\nu = -\frac{2 }{ k} g^{\lambda \mu} (F^\alpha_\mu
F_{\alpha
\nu}-\frac{1}{ 4} g_{\mu \nu} F^{\alpha \beta} F_{\alpha \beta}) \label{enimem}
\ee
The Lagrangian of the Einstein-Maxwell theory is the sum of $L_H$ and
$L_M= \c{L}_M ds$. Accordingly
 Noether currents are simply obtained for this theory as the sum of
 (\ref{Ece}) with the analogous expression for General Relativity (see
\cite{Robutti}).


\subsection{Energy for the Electromagnetic Field in a $(3+1)$ perspective}

We are now going to apply the definition (\ref{ENE}) for the variation of energy
to obtain an expression for the Einstein-Maxwell Lagrangian, writing for
convenience only
the terms regarding the electromagnetic field, since the ones for the
gravitational field are
well known from the previous Section. To obtain the final result we need to integrate the
superpotential and the correction terms $\tau$ on the boundary $B_t$ of the spacetime slice
$\Sigma_t$, which in turn is assumed to be a Cauchy surface. We shall consider a $(3+1)$
splitting of the dynamical fields evolving with respect to boosted observers, which means that we
put ourself in a reference frame which corresponds to observers evolving with respect to $\xi$.\\
We define the electric vector field $E$ and magnetic vector field $B$ in the usual way:
$$
\cases
{E_\mu= F_{\nu \mu } u^\nu\cr
B_\mu=  u^\nu \ep_{\nu  \mu  \alpha \beta}  F^{\alpha \beta}
}
$$
where $ \ep_{\nu  \mu  \alpha \beta}$ is the totally skew-symmetric Levi-Civita tensor on $M$.
Conversely the curvature can be defined in terms of the
electric and the magnetic vector fields as:
\ba
F_{\alpha \beta} = ( E_\alpha  u_\beta-E_\beta u_\alpha )+ u^\nu \ep_{\nu
\alpha \beta \mu} B^\mu
\ea
We define the \textit{electrostatic potential}  and the \textit{normal
component of the electric field} as:
$$
\cases
{\Phi=-A_\alpha u^\alpha \cr
E^\perp=F_{\nu \mu } u^\nu n^\mu=E_\mu n^\mu
}
$$
Analogously, by ($\hat{A}_\mu, \hat{E}_\mu,
\hat{B}_\mu$) we denote  the quantities projected on the $2$-dimensional surface
$B_t$, so that, e.g.:
\be
\hat{A}_\mu=\sigma^\nu_\mu {A}_\nu \Rightarrow {A}_\mu=\hat{A}_\mu+({A}_\nu
n^\nu)n_\mu+\Phi u_\mu
\ee
(and analogous expressions for the other quantities). The symplectic current defined in
(\ref{symp}) owing to (\ref{FemM}) turns out to be:
\be
\omega^\lambda ( L_M, X,\pounds_{\xi}\sigma)=
\frac{2}{ k} [(\delta_X f^{\mu \lambda}) (\pounds_\xi A_\mu)-
(\pounds_\xi f^{\mu \lambda}) (\delta_X A_\mu)   ] \label{symem}
\ee
This expression justifies the definition of $f^{\mu \lambda}$ as the conjugate
momentum with respect to the electromagnetic field.  The pullback of the symplectic current
(\ref{symem}) on the surface $\Sigma_t$ can be written as:
\ba
\omega ( L_{M}, X,\pounds_{\xi}\sigma)&=&[\delta_X {\c{E}}^\mu (\pounds_\xi A_\mu)-\pounds_\xi
{\c{E}}^\mu (\delta_X A_\mu)
] d^3 x=\\
&=&[\delta_X {\c{E}}^\mu (\pounds_\xi A_\mu)-\dot{\c{E}}^\mu (\delta_X A_\mu)] d^3 x \nonumber
\ea
where we have defined the \textit{$3$-dimensional electric density} $\c{E}^\mu$ and
its time rate of change by:
$$
\cases
{\c{E}^\mu=\frac{2}{ k} \sqrt{h} E^\mu\cr
\dot{\c{E}}^\mu=\pounds_\xi {\c{E}}^\mu
}
$$
Using the formula (\ref{LieA}) and recalling that the total order of Maxwell-Einstein
theory is $s=3$, we can
split $\int_\Sigma \omega$, in accordance with (\ref{symp}), into two bulk terms and a boundary
term:
\ba
\int_{\Sigma_t} \omega &=&\int_{\Sigma_t}  \Big{\{} 
\delta_X {\c{E}}^\mu ( \dot{A}_\mu) - \dot{\c{E}}^\mu (\delta_X A_\mu) \Big{\}} d^3 x
\label{omegaA}+\\
&-&q \Big{\{} \int_{\Sigma_t} \delta_X (D_\mu {\c{E}}^\mu) d_\rho \xi^\rho d^3 x
+ \delta_X \int_{B_t} \Big[ \frac{  {\c{E}}^\mu n_\mu}{\sqrt{h}} d_\rho \xi^\rho 
\sqrt{\sigma}  \Big] d^2 x\Big{\}}
\nonumber
\ea
Finally it is possible to identify the reduced  symplectic current and the
term $\tau$ (see (\ref{symp})):
\ba
\tilde{\omega}(L_M, X, \pounds_{\xi}\sigma) &=& [\delta_X {\c{E}}^\mu (
\dot{A}_\mu)-\dot{\c{E}}^\mu (\delta_X A_\mu)] d^3 x \label{pappa}\\
\tau (L_M, X, \pounds_{\xi}\sigma)&=& \frac{2 q}{k}  \delta_X \{ {E}^\mu n_\mu   \sqrt{\sigma}
\}  (d_\rho \xi^\rho) d^2 x
\label{tauem}
\ea
while the remaining term in (\ref{omegaA}) is identically vanishing on shell.
This last expression encodes the symplectic structure of the theory and defines both the phase
space and the conjugate
momenta \cite{Wald}. We remark that the boundary part $\tau (L_M, X, \pounds_{\xi}\sigma)$ of the
symplectic current is non vanishing since this holds whenever the total order of the theory is
higher than
$2$: for electromagnetic theory, treated as a natural theory, the total order is in fact $s=3$.
This term
 $\tau_M$ will influence the definition of the variation of energy in accordance
with (\ref{ENE}).\\
The reduced current (\ref{Ece}) for this theory can be rewritten as:
\ba
\tilde{\c{E}} (L_M, {\xi})=&-&\{  (N \c{T}^\alpha_\nu u_\alpha
u^\nu)+(\c{T}^\lambda_\alpha u_\lambda N^\nu)\} \sqrt{h} d^3 x+ [\c{G}
A_\rho \xi^\rho] \sqrt{h} d^3 x+\\
&+&q [ u_\alpha  J^{(\alpha} \delta^{\nu)}_\sigma \pa_\nu \xi^\sigma-\pa_\nu \big(   J^{[\alpha}
\delta^{\nu]}_\sigma \big) \xi^\sigma u_\alpha]\sqrt{h} d^3 x
\ea
where we have set:
\ba
\c{G}=- {D}_\beta \c{E}^\beta
\ea
which is the spatial part of Gauss' constraint and vanishes on-shell. If we define:
$$
\cases{
\c{H^M}=- \sqrt{h} ( \c{T}^\lambda_\nu u_\lambda u^\nu)=\frac{\sqrt{h}}{
k} (E^\alpha E_\alpha+B^\alpha B_\alpha)\cr
\c{H^M}_\alpha= \sqrt{h} h^\beta_\alpha (\c{T}^\lambda_\beta u_\lambda
)=\frac{2 \sqrt{h}}{ k} (\ep_{\alpha \lambda \beta} E^\lambda B^\beta)
}
$$
we finally  have that:
\ba
\int_{\Sigma_t} \tilde{\c{E}}(L_M, \xi)=\int_{\Sigma_t} [N (\c{H^M}- \Phi \c{G}
)+N^\alpha (\c{H^M}_\alpha+A_\alpha \c{G})] d^3 x+ \label{etildem} \\
+ q \int_{\Sigma_t}
[u_\alpha  J^{(\alpha} \delta^{\nu)}_\sigma \pa_\nu \xi^\sigma-\pa_\nu \big(   J^{[\alpha}
\delta^{\nu]}_\sigma \big) \xi^\sigma u_\alpha] \sqrt{h}
d^3 x \nonumber
\ea
To calculate the variation of the energy by means of (\ref{ENE})  we calculate now the expression
 for the superpotential, evaluated on $B_t$:
\ba
\int_{B_t} \c{U} (L_M, \xi ) d^2 x&=& \frac{2 }{ k}\int_{B_t}
E_\perp [-N \Phi+\hat{A}_\rho {N}^\rho +v N ({A}_\nu n^\nu)] \sqrt{\sigma} d^2
x+ \label{uemtre}\\
&-&q \delta_X \int_{B_t} J^{[\alpha} \delta^{\nu]}_\sigma
{u}_\nu {n}_\alpha \xi^\sigma \frac{\sqrt{\sigma} }{\sqrt{g}} d^2
x¯\ +\frac{2 q}{ k} \delta_X \int_{B_t}  \{ {E}^\mu n_\mu  d_\rho \xi^\rho
\sqrt{\sigma}
\} d^2x \nonumber
\ea
together with the correction term, which turns out to be:
\ba
\int_{B_t} i_\xi \Bbb{F} (L_M,X)=\frac{2}{ k} \int_{B_t} N( E_\perp u^\mu
+ v E^\mu+u_\delta n_\nu \ep^{ \delta  \nu \mu \beta} B_\beta )
\delta_X A_\mu
\sqrt{\sigma} d^2 x \label{correm}
\ea
Neglecting terms which vanish on shell, from (\ref{tauem}), (\ref{uemtre}) and (\ref{correm})  we
obtain that the variation of energy for the electromagnetic part of the Einstein-Maxwell theory
is:
\ba
\delta_X E (L_M, \xi)& =& \int_{B_t} [\delta_X \c{U} -i_\xi
\Bbb{F}-\tau] d^2 x=
\label{deem}\\ &=&-\frac{2 }{ k} \int_{B_t} N[ \Phi-v (n^\alpha A_\alpha)] \delta_X
(E^\perp \sqrt{\sigma}) d^2 x+\frac{2 }{ k} \int_{B_t} {N}^\alpha  \delta_X
(E^\perp  \hat{A}_\alpha \sqrt{\sigma})d^2 x + \nonumber \\
&-&\frac{2 }{ k} \int_{B_t} N [v \hat{E}^\alpha +  u_\nu n_\beta
\ep^{\nu   \beta  \alpha \mu} \hat{B}_\mu ] \delta_X \hat{A}_\alpha
\sqrt{\sigma} d^2 x \nonumber
\ea
We will see that the first term in the last  expression is directly related with
the charge of the
field, the second  is related with the rotational degrees of freedom and
the third one is
determined
by the values of the electric and the magnatic field projected onto  the surface $B_t$. \\
We remark that this expression of energy exactly cancels the boundary terms
in the variation of  the reduced current (\ref{etildem}) so that, in accordance with the general
theory, we obtain that the variation of the Hamiltonian (\ref{varacca})
\be
\delta_X {H} (L_M, \xi, \Sigma )=\int_{ \Sigma}  \delta_X \tilde{\c{E}}  (L_M, \xi, \sigma )+
\int_{\pa \Sigma} [ \delta_X \c{U} (L_M, \xi )-i_\xi \Bbb{F} (L_M,X)-\tau ( L_M ,
X,\pounds_{\xi}\sigma) ]
\ee
is a pure bulk term, which, according to (\ref{etilde}) and (\ref{pappa}), generates the
Hamiltonian equations of motion.\\


\subsection{Energy and boundary conditions}

In the Einstein-Maxwell theory we have that Hilbert Lagrangian is minimally coupled
with the Maxwell Lagrangian, so that the conserved quantities are additive:
\be
\delta E( L_{EM}, \xi)= \delta E ( L_{H}, \xi)+\delta E ( L_{M}, \xi)
\ee
We have previously defined in (\ref{E-EO}) the quasilocal internal energy for the gravitational
field contained in a region of spacetime bounded by a surface $\c{B}$, imposing the Dirichlet
metric boundary conditions (\ref{Diri}).
In analogy, we want to derive now the energy contribution due to the pure
electromagnetic part
of the Einstein-Maxwell theory. To perform calculations we rewrite equation 
(\ref{deem}) in terms of
boosted observers:
\ba
\delta_X E( L_{M}, \xi) &=&\label{puioy}-\frac{2 }{ k} \int_{B_t} \bar{N}  \bar{\Phi}
\delta_X (\bar{E}^\perp \sqrt{\sigma}) d^2 x +\frac{2 }{ k} \int_{B_t} \bar{N}^\alpha  \delta_X
(\bar{E}^\perp  \hat{A}_\alpha \sqrt{\sigma})d^2 x \label{vareelem}\\
&-&\frac{2 }{k} \int_{B_t} \bar{N}  [  \bar{u}_\nu \bar{n}_\beta
\ep^{\nu   \beta  \alpha \mu} \hat{\bar{B}}_\mu \delta_X \hat{A}_\alpha ]
\sqrt{\sigma} d^2 x \nonumber
\ea
where we recall from from the definitions given above that $E^\perp=\bar{E}^\perp$,
$\bar{\Phi}=-{A}_\alpha
\bar{u}^\alpha$,
$\bar{B}=\bar{u}^\nu \ep_{\nu  \mu  \alpha \beta}  F^{\alpha \beta}$ and $ \bar{u}_\nu
\bar{n}_\beta \ep^{\nu   \beta  \alpha \mu}= \ep^{\nu   \beta  \alpha \mu} {u}_\nu {n}_\beta$.\\
To integrate this variational equation and consequently to obtain the
energy contribution in the region bounded by $B_t$ we have to consider a $1$-parameter family of
solutions of field equations which admit the same physical data fixed on the boundary. We can
choose different control modes which correspond to two different physical situations, namely a
system which is electrically isolated from outside or an adiabatic system (see
\cite{Kijowski}). \\
\begin{itemize}
\item We can choose a control mode for the boundary components of the
electromagnetic
potential $\hat{A}_\alpha$ (i.e. we control the magnetic flux through $B_t$) and for the
electrostatic
potential $\bar{\Phi}$, which is equivalent to state that the laboratory is electrically
isolated from outside. In other words we assume $\delta_X \hat{A}_\alpha\mid_{B_t }=0$ and
$\delta_X
\bar{\Phi} \mid_{B_t }=0$ in (\ref{puioy}).  Recalling that Dirichlet conditions (\ref{Diri}) are
chosen for the gravitational field $\delta_X \bar{\gamma} \mid_{\c{B}}=0$, we obtain from
(\ref{vareelem}) that:
\be
\delta_X E (L_M, \xi) =
 \delta_X \Big{\{} -\frac{2 }{ k} \int_{B_t} \bar{N}  \bar{\Phi}
\bar{E}^\perp \sqrt{\sigma} d^2 x +\frac{2 }{ k}  \int_{B_t} (\bar{N}^\alpha
\bar{E}^\perp  \hat{A}_\alpha \sqrt{\sigma})d^2 x \Big{\}}
\ee
This expression is integrable so that:
\ba
 E (L_{M}, \xi) -{E_0} (L_{M}, \xi)&=&-\frac{2 }{ k} \int_{B_t} (\bar{N}
\bar{\Phi}
-\bar{N}^\alpha  \hat{A}_\alpha ) \bar{E}^\perp \sqrt{\sigma} d^2 x \label{Eemeo}=\\
&=&\frac{2 }{ k} \int_{B_t} (\xi^\mu {A}_\mu)  \bar{E}^\perp \sqrt{\sigma} d^2 x \nonumber
\ea
We recall that the variation is performed along a one parameter family of
solutions, all admitting  the same boundary conditions. Hence $E_0 $ is the energy
corresponding to a reference solution inside this family.\\ Expression (\ref{Eemeo}) for the
energy of the electromagnetic field, owing the boundary
conditions for the gravitational and the electromagnetic field ($\bar{\Phi}
\mid_{B_t}=\bar{\Phi}_0 \mid_{B_t}$, ${{\hat{A_0}}}_\alpha
\mid_{B_t}={\hat{A}}_\alpha
\mid_{B_t}$, $\bar{\gamma}_{\mu \nu}={\bar{\gamma_0}}_{\mu \nu}$), can be rewritten in an
analogous way as:
\be
 E (L_M, \xi)=-\frac{2 }{ k} \int_{B_t} (\bar{N}  \bar{\Phi}
-\bar{N}^\alpha  \hat{A}_\alpha ) (\bar{E}^\perp-\bar{E}_0^\perp) \sqrt{\sigma} d^2 x
\ee
where the subscript $0$ clearly refers to the reference solution. This formula reproduces the
energy content of the electromagnetic field and it stresses that
the response variable, with this boundary condition, is the normal component of
the electric field. The one parameter curve of solutions is parametrized by $\bar{E}^\perp$, i.e.
by its electric charge.

\item If we choose instead a control mode for $\hat{A}_\alpha$ and $\bar{E}^\perp$, (i.e.
$\delta_X
\hat{A}_\alpha\mid_{B_t }=0$ and $\delta_X
 \bar{E}^\perp\mid_{B_t }=0$) we
state that there
is no flux of  magnetic and electric field through $B_t$. In analogy with classical
thermodynamics this system can be defined to be \textit{adiabatic}. In this case calculations are
even more trivial. From equation (\ref{puioy}) and always assuming Dirichlet boundary conditions
(\ref{Diri}) we find:
\be
\delta E (L_M, \xi)=0 \Rightarrow  E (L_M, \xi)=\text{const}
\ee
This result, which appears to be quite shocking, can be interpreted by saying
that the contribution to energy coming from the pure electromagnetic Lagrangian is neglectable.
This is in accordance with the fact that the system is adiabatic.  Nevertheless the pure
gravitational contribution (\ref{E-EO}) to the total energy keeps track of the electromagnetic
field through the metric solution $g$ of Einstein equations
$G_{\mu \nu}=k \c{T}_{\mu \nu}$, where $\c{T}_{\mu \nu} $ is the electromagnetic stress tensor
defined in (\ref{enimem}).  In other words, the quasilocal energy (\ref{E-EO}) of the
gravitational field
  takes account of the presence of the electromagnetic field.
\end{itemize}
These results  are in accordance with the analysis on control modes proposed by Kijowski in
\cite{Kijowski}, using a symplectic type analysis in the framework of Legendre transformation and 
Hamiltonian formalism. We remark that formula  (\ref{Eemeo}) generalizes this analysis to the case
of Einstein-Maxwell theory in a spacetime region with non-orthogonal boundaries.


\section{Rigidly rotating horizon}

In this Section we are going to recall the definitions of \textit{isolated horizon},
\textit{weakly isolated horizon} and \textit{rigidly rotating horizon}. The
concept of
isolated
horizon was first introduced by A. Ashtekar and coworkers in \cite{Ashtekar1}
and later
developed in \cite{Ashtekar4} to obtain a viable and more general definition.
This new concept allows to generalize the laws of black holes thermodynamics
to the case of black holes which do not admit an event horizon and a global Killing
vector field. This situation is much more realistic with respect to the
quasi-static laws of black
holes mechanics based on the existence of an event horizon and referred to equilibrium
situations and small perturbations out of them. The definition of isolated horizons is given
intrinsecally and this implies that we do not need to know the whole hystory of spacetime to
define this geometric surface (as we usually need for the event horizon in non stationary
spacetimes). We can admit isolated
horizons with radiation infinitesimally near the horizon surface,  meaning
non-stationary spacetimes. Conditions are imposed only on the internal geometry
of the horizon, not on the whole spacetime, to ensure that the black hole
itself
is "isolated". This is reasonable also in analogy with the classical approach to
thermodynamics \cite{Caratheodory} where equilibrium is just defined in function
of the internal degrees of freedom of the system.\\
Physical examples of isolated horizons can be found in collapsing stars or in
cosmological horizons in de-Sitter spacetime (see \cite{Ashtekar2} for a review on the matter); in
the latter case no singularities are present in spacetime, but it is still possible to define
thermodynamics for such surfaces.\\
The matter covered in this subsequent Section is (apart from minor changes
needed to adapt to our notations) a brief account of Ashtekar's definitions of  \cite{Ashtekar4},
which we consider worth of being recalled for the sake of completeness and clarity.

\subsection{ Geometry of horizons}

The fundamental requirement is to ask the isolated horizon $\Delta$ (a full
definition will be given later) to be a null non expanding $3$-surface, which means that the
$3$-metric on it admits a Killing vector;
this encloses the fundamental concept that
$\Delta$ is isolated in a suitable physical sense. If any matter or
electromagnetic field is in interaction with the gravitational field we have to
state that
the flux of radiation and matter through the horizon is zero, which for the
electromagnetic field is equivalent to say that the relevant component of the
Poynting vector is zero on  $\Delta$. With more precision we can start with the following
definition:
\begin{definition}
A null hypersurface $\Delta$ in a $4$-dimensional spacetime $M$ is a $3$-dimensional submanifold
of the spacetime  admitting a null normal vector $l=l^\mu \pa_\mu$ (which for
definition is also  tangent to the same surface $\Delta$).
\end{definition}
In the sequel we will consider only null surfaces which are diffeomorphic
to the product $\Delta \simeq \R \times S^2$ such that for each $t \in \R$ there
exists a diffeomorphism $\psi_t: S^2 \rightarrow \Delta _t$, where $\Delta
_t$ is a
two dimensional  spacelike surface.\\
\noindent The metric defined on $\Delta$ is clearly degenerate for definition (see
\cite{Waldbook}). We now consider a region $D$ of  spacetime  such that $\Delta \subset D$;
like in the previous Section, we assume $D$ to be foliated by  $3$-dimensional spacelike
hypersurfaces
$\Sigma_t$, which induce a preferred foliation $\hat{\psi}$ of $\Delta$ and we
denote again by $u^\mu$  the future directed causal
normal to $\Sigma_t$  and by $n^\mu$ the outward pointing normal to $\Delta_t$ in
$\Sigma_t$. \\
To define the metric on $\Delta$ we have to introduce a null vector field
$\hat{l}$ transverse to $\Delta$. We ask this vector field to
be normalized relatively to $l$ so that $\hat{l}^\alpha l_\alpha=-1$. Notice that $l^\alpha \in
T \Delta$, $\hat{l}_\alpha \in
T^* \Delta$, while transversality requires  $l_\alpha \notin
T^* \Delta$ and $\hat{l}^\alpha \notin T \Delta$. Due to this normalization condition it is
possible to express $l$ and $\hat{l}$ in terms of
$u^\mu$ and $n^\mu$ as:

\baa
l_\alpha=f (u_\alpha + n_\alpha)\\
\hat{l}_\alpha=\frac{1}{2 f} (u_\alpha-n_\alpha)
\eaa
where $f$ is a non-vanishing function on $M$. Now the "metric" $q_{\alpha \beta}$ and the
projector operator
$q^{\nu}_\beta$ onto $\Delta $ can be defined as:
$$
\cases
{
q_{\alpha \beta}=g_{\alpha \beta}+2 l_{(\alpha} \hat{l}_{\beta)}\cr
q^{\nu}_\beta=g^{\nu \alpha} q_{\alpha \beta}
}
$$
Since $q_{\alpha \beta}$ is degenerate the "inverse" $q^{\alpha \beta}$ of this  "metric" is
not unique but any reasonable definition of it  remains unchanged under the
addition of terms of the form $l^{(\alpha } V^{ \beta)}$, with $V$ tangent to $\Delta$.\\
The non-degenerate metric $\sigma$ over $\Delta_t$ can be obtained by considering
either $\Delta_t$ as a submanifold of $\Delta$ or a submanifold of $\Sigma_t$. We have:
$$
\cases
{
\sigma_{\alpha \beta}=g_{\alpha \beta}+2 l_{(\alpha} \hat{l}_{\beta)}=
g_{\alpha \beta}- n_{\alpha} n_{\beta}+u_{\alpha} u_{\beta}\cr
\sigma^{\nu}_\beta= \delta^{\nu}_\beta+l^\nu \hat{l}_{\beta}+\hat{l}^\nu
{l}_{\beta}
}
$$
where the analytical expression of $\sigma_{\alpha \beta}$ is the same of $q_{\alpha \beta}$, but
has not to be confused with $q_{\alpha \beta}$ which is defined in a different space. The time
evolution, defined by a vector field
$\xi^\alpha$ transverse to
$\Sigma_t$, see  (\ref{tempo}), can be expressed as an element of $T\Delta$ as:
\ba
\xi^\alpha=\tilde{N} l^\alpha+\bar{N}^\alpha \label{put}
\ea
where $\tilde{N}=\frac{N}{f}$ and $\bar{N}^\alpha=\sigma^\alpha_\beta N^\beta$.\\ For each null
hypersurface
$\Delta$ it is possible to select (non uniquely) a vector $l^\mu$  (which is for definition a
null normal vector) and define the
\textit{tensor expansion} of $\Delta $ as:
\be
{\theta_{(l)}}_{\alpha \beta}=q^{\nu}_\alpha q^{\mu}_\beta  \n_\nu l_\mu \label{pit}
\ee
The expansion plays for null surfaces the same role that extrinsic
curvature plays
for spacelike or timelike hypersurfaces. As $l^\alpha$ is normal to $\Delta$ we have
immediately, from  Frobenius' theorem, that the twist of the expansion is zero
${\theta_{(l)}}_{[ \alpha \beta ]}=0$, so that $\theta_{ \alpha \beta }$ can be
decomposed
as:
\be
{\theta_{(l)}}_{\alpha \beta}=\frac{1}{2}\theta_{(l)} \sigma_{\alpha
\beta}+\zeta_{\alpha
\beta}
\ee
where $\zeta_{\alpha \beta}=\zeta_{(\alpha \beta)}$ is called the \textit{shear} of the expansion
tensor. The
scalar: 
\be
\theta_{(l)}={\theta_{(l)}}_{\alpha \beta} \sigma^{\alpha \beta}= \frac{1}{\sqrt{\sigma}}
\pounds_l (\sqrt{\sigma})
\ee
is properly called \textit{expansion} of the null surface (see
\cite{Waldbook}). We are now ready to give the definition \cite{Ashtekar1}:

\begin{definition} \label{1}
A non-expanding horizon $\Delta \subset M$ is a $3$-dimensional submanifold of
spacetime $(M,g)$

with the following properties:

\begin{itemize}

\item $\Delta $ is a null hypersurface and it is topologically the product of
a two-dimensional sphere and the real line: $\Delta \cong S^2 \times \R$

\item The expansion $\theta_{(l)}$ of any null normal to the surface vanishes
on $\Delta$

\item Equations of motion hold at $\Delta $ and the stress energy tensor
for matter interacting with gravity is such that $(-T_{\alpha \beta}l^\alpha )\mid_\Delta$ is
future directed and causal for each future directed null normal $l$ to $\Delta $
\end{itemize}
\end{definition}
We remark that the fundamental property to identify an isolated
horizon is to impose $\theta_{(l)}=0$. The topological condition is equivalent to the
requirement that  there exists a foliation of the hypersurface $\Delta$ in "time" constant
surfaces. The matter condition is very weak and it is implied for example
from the
dominant energy condition; it is equivalent to state that there is no flux of
matter through the horizon; see \cite{Ashtekar1} for a detailed presentation.\\
Since the twist of $\theta_{(l)}$ vanishes, from the Raychaudhuri equation \cite{Raychaudhuri},
we obtain that every null
normal $l$ is free of expansion, twist and shear on the non-expanding horizon, i.e.
${\theta_{(l)}}_{[\alpha \beta]} \mid_\Delta=0 $, $\theta_{(l)}\mid_\Delta =0$ and
$\zeta_{\alpha
\beta}\mid_\Delta=0$.\\  The same Raychaudhuri equation ensures that there are some
restrictions on the Ricci tensor and in particular that the following should hold
\baa
R_{\alpha \beta} l^{\alpha } l^{ \beta} \mid_\Delta=0
\eaa
The \textit{natural connection $1$-form} ${\omega}$ is defined on $\Delta$ 
(since $l$ is expansion and twist free) as:
\ba
 q^\mu_\alpha \n_\mu l^\beta =(\omega_\alpha  l^\beta) \label{omegadef}
\ea
which implies the property for $l$ to be a Killing vector of the $3$-metric $q$:
\baa
\pounds_l q_{\alpha \beta}=0
\eaa
The \textit{acceleration} $k_{(l)}$ of $l$, defined as:
\ba
l^\mu \n_\mu
l^\beta=k_{(l)} l^\beta \label{kappaelle}
\ea
is expressed in terms of the natural connection as $k_{(l)}=\omega_\alpha l^\alpha$
\cite{Ashtekar1}. Notice that
$k_{(l)}$ is the counterpart for isolated horizons of the surface gravity for Killing horizons
\cite{Waldbook}.\\  Finally we stress that from the third condition in Definition $2$ it follows
that the stress energy tensor contracted twice with $l^\alpha$ is identically zero, i.e. 
$T_{\alpha
\beta}l^\alpha l^\beta=0$. This implies that in the case of Einstein-Maxwell theory we are
dealing with, 
$f_\mu=F_{\alpha
\beta}l^\alpha \sigma^\beta_\mu=0$ and $\hat{f}_\mu=F_{\alpha \beta}
\hat{l}^\alpha \sigma^\beta_\mu=0$. This is equivalent to state that the
electric and the magnetic components tangential to $\Delta_t$ are identically zero.
\begin{definition}
\label{2}
A non-expanding horizon becomes a \textit{weakly isolated horizon} $(\Delta,
[l])$  if it is equipped with an equivalence class $[l]$  of null normals satisfying 
the equation:
\be
\pounds_l \omega=0, \forall l \in [l] \label{gaugeom}
\ee
where the equivalence relation is defined as: $l \sim l' \iff l=c
l'$; $c \in \R $ is a non zero constant.
\end{definition}
We remark that a Killing horizon is automatically a weakly isolated horizon. As it happens for
Killing horizons it is possible to define (apart from a constant factor, which depends on the
representative inside $[l]$) the surface gravity for weakly isolated
horizons if we equip them (as it has been done in the definition) with a
preferred family of null normals. The value of $k_{(l)}$ depends on the element of
the class we choose. One can otherwise uniquely define $k_{(l)}$ by demanding it to be a
suitably defined function on the horizon; see \cite{Ashtekar2}.\\
The properties of weakly isolated horizons ensure that the \textit{zero
law} holds true.
This is
equivalent to say that the surface gravity is constant on a weakly isolated
horizon. In
fact
from the property $\pounds_l \omega=0$, we obtain that:
\be
d (k_{(l)}) \mid_\Delta=0
\ee
We will consider for applications only the case of rigidly rotating
horizons interacting with electromagnetic fields, i.e. the Einstein-Maxwell described above. The
theory developed hereafter is also true in the case of isolated horizons, which can be considered
as particular cases of rigidly rotating horizons (roughly speaking with vanishing angular
momentum).
\footnote{
We will not use the above definition in the sequel, but we report it for
completeness:
\begin{definition} \label{3}
A weakly isolated horizon becomes an \textit{isolated
horizon}
$(\Delta, [l])$ if it is  equipped with an equivalence class of null normals to
it satisfing the equation:
\be
[\pounds_l, D] V=0
\ee
for any vector field $V$ tangential to $\Delta$ and $l \in [l]$, where $D$
denotes
the covariant derivative with respect to the metric on $\Delta$.
\end{definition}
We stress however that properties holding for weakly isolated
horizons are also true for isolated horizons. In fact it is easy to show that every isolated
horizon is a weakly isolated horizon (see for details \cite{Ashtekar2}).
}

\begin{definition} \label{4}
A weakly isolated horizon $(\Delta,[l])$ is said to be a rigidly
rotating horizon $(\Delta, [l],\varphi)$  if it admits a rotational
simmetry $\varphi$ (with $\varphi$ tangent to the surfaces 
$\Delta$) with closed and circular orbits, such that $\pounds_\varphi
l^\alpha=0$, $\pounds_\varphi
\omega_\alpha=0$ and $\pounds_\varphi q_{\alpha \beta}=0$.
\end{definition}
This last definition is a restriction on the class of weakly isolated
horizons we are considering
and it is analogous to state that the $3$-dimensional geometry of the
hypersurface is
axisymmetric,
with $\varphi$ as its infinitesimal rotational symmetry generator \cite{Ashtekar4}.\\
In this case, it is possible to choose a foliation of $\Delta$ in a preferred way so
that the surfaces
$\Sigma_t$ intersect $\Delta$ at the preferred  $2$-surfaces $\Delta_t$
topologically diffeomorphic to $S^2$ and we state that there exists a tangent
vector $\varphi^\alpha \in  T \Delta_t$, which generates the desired symmetry.
Correspondingly,
the time flow transverse to the surface  $\Sigma_t$ is adapted to $\Delta$ if
$\xi^\alpha= T^\alpha= l^\alpha-\Omega_{(l)}  \varphi^\alpha$ on $\Delta$.\\
In the case of interaction with matter, and particulary in the case of
Einstein-Maxwell theory
we are going to analyse, it is natural to require $\varphi$ to be a
symmetry also for the electromagnetic field projected on $\Delta$, so that $\pounds_\varphi
(q^\alpha_\beta q^\mu_\nu F_{\alpha \mu})=0$.\\
In stationary and asymptotically flat spacetimes one usually sets the electrostatic potential
to be $\Phi=-\xi^\mu A_\mu$ and requires a gauge fixing on the electromagnetic potential such that
$A_\mu$ tends to zero at infinity and $ \pounds_\xi A_\mu=0$ everywhere in
spacetime. In analogy with this fact we define a \textit{gauge fixing adapted to the
weakly isolated horizon} by setting:
\be
\pounds_l A_\mu \mid_\Delta =0
\ee
This choice of the gauge is only formally analogous to the condition (\ref{gaugeom}) imposed on
$\omega$ (i.e. $\pounds_l \omega=0 $) because this latter is a restriction on the form of the
gravitational field, while the gauge fixing can be imposed without
constraining the electromagnetic field itself.  
We can now define the horizon \textit{electrostatic potential}  as:
\be
\Phi_{(l)}=-l^\mu A_\mu
\ee
and the gauge fixing condition ensures that $\Phi_{(l)}$ is constant on the
horizon \cite{Ashtekar1}
\be
\pounds_l A_\mu \mid_\Delta=0 \Rightarrow d(\Phi_{(l)})\mid_\Delta =0
\ee
This property, together with the zero principle and the boundary conditions
which define the
geometry
of a rigidly rotating horizon, will ensure that the parameters $\Phi_{(l)}, k_{(l)},
\Omega_{(l)}$, which appear in the first principle of thermodynamics for rigidly rotating
horizons, are constant on $\Delta$.


\subsection{Energy for rigidly rotating horizon}
The formalism developed for calculating the variation of energy of the
gravitational and the electromagnetic field naturally applies to define the
variation of
energy  of a region of spacetime bounded by an isolated horizon.\\
We do not mind in the sequel  what
happens on the outer boundary $\c{B}$; this case has been deeply examined in
\cite{raiteri}
 to which we refer the reader for details.  We rewrite the expression for energy variation
in the case of Einstein-Maxwell theory, using a $2$-surface with no a priori defined geometry.
Gluing together formula (\ref{degr}) and formula (\ref{puioy}), we find:
\ba
\delta E (L_{EM}, \xi)&=& \int_{B_t} \Big{\{}  \bar{N} \delta_X ( \sqrt{\sigma}
\bar{\epsilon})
-\bar{N}^\alpha \delta_X (\sqrt{\sigma} \bar{j_\alpha})+\frac{\bar{N}
\sqrt{\sigma}}{ 2}
\bar{s}^{\mu \nu} \delta_X \sigma_{\mu \nu}\Big{\}} d^2 x+ \label{patata} \\
&+& \frac{1}{k}\int_{B_t} [\pounds_\xi
(\sqrt{\sigma}) \delta_X (\theta)- \delta_X (\sqrt{\sigma})
\pounds_\xi(\theta)]  d^2 x \nonumber\\
&-&\frac{2 }{ k} \int_{B_t} \bar{N} \bar{\Phi} \delta_X
(\bar{E}^\perp \sqrt{\sigma}) d^2 x +\frac{2 }{ k} \int_{B_t} \bar{N}^\alpha  \delta_X
(\bar{E}^\perp  \hat{A}_\alpha \sqrt{\sigma})d^2 x+  \nonumber \\
&-&\frac{2 }{ k} \int_{B_t} \bar{N}  [ \bar{u}_\nu \bar{n}_\beta
\ep^{\nu   \beta  \alpha \mu} \hat{\bar{B}}_\mu \delta_X \hat{A}_\alpha]
\sqrt{\sigma} d^2 x  \nonumber
\ea
This formula expresses the variation of energy for the Einstein-Maxwell theory
 evaluted on a generic surface $B_t$. From now we shall specialize this formula to the case of
rigidly rotating horizons, i.e. to the case of $B_t = \Delta_t$.  We stress that
in that case we chose a null evolution of the boundary and this implies
that the boost velocity, defined in (\ref{velboost}), is $v=1$. \\
Using the  boost relations (see \cite{BLY}), we obtain:
$$
\cases
{\bar{\epsilon}=\frac{1}{ k} [\gamma \c{K} +\gamma v l] \cr
\bar{j_\alpha}=\frac{1}{ k} \sigma_\alpha^\beta n^\gamma K_{\beta \gamma}-\frac{1}{ k}
\n_\alpha
\theta=j_\alpha-\frac{1}{ k}
\n_\alpha
\theta\cr
\bar{n}^{\alpha} \bar{a}_{\alpha}= \gamma n^\alpha a_\alpha- \gamma v u^\alpha  b_\alpha +
\bar{u}^\alpha
\n_\alpha
\theta }
$$
where we have defined $l_{\mu \nu}= -\sigma^\alpha_\mu  \sigma^\beta_\nu \n_\alpha u_\beta$, so
that $l=l_{\mu \nu} \sigma^{\mu \nu}$ and $b^\nu=n^\alpha \n_\alpha n^\nu$. \\
After a long calculation (details can be found in \cite{BLY}, \cite{raiteri} for the
gravitational part, while for the electromagnetic part we refer to formula (\ref{deem})), the
equation which defines the variation of energy on a generic null surface can be rewritten,
starting from (\ref{patata}) ,  as:
\ba
\delta E_{\Delta}(L_{EM}, \xi)&=&
-\frac{1}{k}\int_{\Delta_t} \Big{\{}  \tilde{N}\delta_X \big( \sqrt{\sigma}{\theta_{(l)}} \big)
-\bar{N}^\alpha \delta_X \big( \sqrt{\sigma} {\hat{\omega}_\alpha}
\big)+\frac{\tilde{N} \sqrt{\sigma}}{ 2}
{{s}_\Delta}^{\mu \nu} \delta_X \sigma_{\mu \nu} \Big{\}}d^2 x+\nonumber \\
&-&\frac{2 }{ k} \int_{\Delta_t} \tilde{N}  {\Phi}_{(l)} \delta_X
(E^\perp \sqrt{\sigma}) d^2 x +\frac{2 }{ k} \int_{\Delta_t} \bar{N}^\alpha  \delta_X
(E^\perp  \hat{A}_\alpha \sqrt{\sigma})d^2 x \label{ERRH}\\
&-&\frac{2 }{k} \int_{\Delta_t} \tilde{N}   f^\alpha \delta_X \hat{A}_\alpha
\sqrt{\sigma} d^2
x+ \int_{\Delta_t} \{  (\pounds_\xi P_\Delta^{\sqrt{\sigma}} ) \delta_X
\sqrt{\sigma}-(\pounds_\xi \sqrt{\sigma}) \delta_X P_\Delta^{\sqrt{\sigma}}
\} d^2 x\nonumber
\ea
where we have set:
$$
\cases{
{s}_\Delta^{\alpha \beta}= [ {\theta_{(l)}}^{\alpha
\beta}-(k_{(l)}+{\theta_{(l)}}) \sigma^{\alpha \beta} ]\cr
P_\Delta^{\sqrt{\sigma}}=k \ln (f)
}
$$
The acceleration $k_{(l)}$, in accordance with definition (\ref{kappaelle}), is expressed by:
\ba
k_{(l)}=f [ n^\alpha a_\alpha- u^\alpha b_\alpha ]+ k \pounds_l (P_\Delta^{\sqrt{\sigma}})
\ea
and the natural connection $1$-form $\hat{\omega}_\alpha=\sigma_\alpha^\beta {\omega}_\beta$, in
accordance with definition (\ref{omegadef}), turns out to be:
\ba
\hat{\omega}_\alpha=\sigma_\alpha^\beta {\omega}_\beta=\sigma_\alpha^\beta n^\gamma \n_\beta
u_\gamma+k d_\alpha P_\Delta^{\sqrt{\sigma}}=-k j_\alpha+k d_\alpha P_\Delta^{\sqrt{\sigma}}
\ea
To perform the calculations from (\ref{patata}) to (\ref{ERRH})  we also made use of the following
relations, which can be easily obtained from (\ref{put}) and (\ref{pit}):
$$
\cases{
 {\theta_{(l)}}_{\alpha \beta}=- \xi (\c{K}_{\alpha \beta}+ l_{\alpha \beta})\cr
\pounds_\xi \sqrt{\sigma}=\tilde{N}  \sqrt{\sigma} {\theta_{(l)}}+\sqrt{\sigma} d_\beta
\bar{N}^\beta
 }
$$
Up to now formula (\ref{ERRH}) holds true for any cross section $\Delta_t$ of a generic null
hypersurface $\Delta$. In the case of a rigidly rotating horizon it is possible to obtain
explicitly a first principle of thermodynamics if we recall the
more relevant geometric properties we explained in section (5.1), which hold true on that
particular null surface:
\begin{itemize}
\item ${\theta_{(l)}} \mid_\Delta=0$
\item $k_{(l)}$ and $\Phi_{(l)}$ are constant on $\Delta$
\item $f_\mu=F_{\alpha \beta}l^\alpha \sigma^\beta_\mu=0$ and $\hat{f}_\mu=F_{\alpha
\beta}\tilde{l}^\alpha
\sigma^\beta_\mu=0$
\item $\xi^\alpha=T^\alpha=l^\alpha- \Omega_{(l)} \varphi^\alpha$ on $\Delta$
\item $\varphi$ is a symmetry for both the $3$-metric and the electromagnetic field
\end{itemize}
With these assumptions, which  follow directly  from the definition of rigidly
rotating horizon, it is  straightforward to rewrite  (\ref{ERRH}) as:
\be
\delta E_{\Delta}(L_{EM}, \xi)=\frac{k_{(l)}}{k} \delta_X A_\Delta+
\Phi_{(l)} \delta_X Q_\Delta+
\Omega_{(l)} \delta_X J_\Delta \label{fprrh}
\ee
where we define the area, the  charge and  the angular momentum of the horizon as follows:
$$
\cases
{
A_\Delta=\int_{\Delta_t}  \sqrt{\sigma} d^2 x \cr
Q_\Delta=-\frac{2}{k}\int_{\Delta_t} E_\perp \sqrt{\sigma} d^2 x \cr
J_\Delta= -\frac{1}{k} \int_{\Delta_t} \varphi^\alpha (\hat{\omega}_\alpha +2 E_\perp
\hat{A}_\alpha) \sqrt{\sigma} d^2 x
}
$$
and owing to the definition of rigidly rotating horizon the quantities $A_\Delta$ and $Q_\Delta$
are independent on time, i.e. on the slice $\Delta_t$ we are integrating on. The definition of
rigidly rotating horizon ensures that also  $J_\Delta$ is independent on the slice $\Delta_t$ we
are integrating on; see   \cite{Ashtekar4} and \cite{BoothIH} for a detailed discussion.
\\ The integrability of this expression depends on the calibration of the
parameters (namely
the "temperature" $=\frac{k_{(l)}}{k}$, the angular velocity $\Omega_{(l)}$ and the
electrostatic potential $\Phi_{(l)}$). This argument are
extensively treated in other papers and we refer  for details to \cite{BoothIH},
\cite{Ashtekar4}. We just want to point out that the Noether covariant approach together with a
proper analysis of boundary terms in the definition of energy allows us to reproduce the first
law of thermodynamics also in the case of   rigidly rotating horizons (\ref{fprrh}) in exactly
the same way as it was formulated in
\cite{Ashtekar4}.

\section{Conclusions}
Our analysis allowed us to formulate a first principle for rigidly
rotating horizons analogous
to the ones given in  \cite{Ashtekar4} and \cite{BoothIH},
but the approach and the formalism used here are completely different. The approach used in
 \cite{Ashtekar4} is metric-affine, based on tetrad gravity, and uses a symplectic type analysis
to define the conserved quantities. On the other hand, Booth  uses in \cite{BoothIH}  a purely
metric approach based on a canonical decomposition of a "modified" trace-K action functional.
Indeed he needs to add suitable boundary terms in the action functional to correct the conserved
quantities. \\ 
We worked instead in a purely metric framework and we considered a definition of the conserved
quantities which naturally arises from the Noether theorem. Closed in spirit with the methods
employed in the covariant ADM approach (\cite{CADM}, \cite{Sinicco}) and the Regge-Teitelboim
analysis of boundary terms in the Hamiltonian variation, we are able to correctly define the
variation of the Hamiltonian and the variation of energy by pushing all the boundary terms
arising in the variation of the Noether current into the definition itself. \\ This allowed us to
treat succesfully Einstein-Maxwell theory, where Electromagnetism is considered as a natural
theory. The framework allows us to define the theory from a
geometric point of view, without a priori assumptions on the magnetic
field and on the
existence of
a global potential. Energy definitions arising from the imposed boundary
conditions have a
nice physical interpretation.\\
Finally this theory find  applications in defining a first
law for rigidly rotating horizons, which have recently assumed  importance in trying to
generalize the quasi static treatments of thermodynamics of black
holes.  These surfaces allow us to  define the first
principle also for non-static solutions, that admit radiation near the
horizon; quasilocal formalism is indispensable in this context.\\
We have treated here the case of minimal interaction between the
gravitational field and the electromagnetic radiation using Einstein-Maxwell Lagrangian. It is
straightforward to include into the theory a dilatonic coupling with the Maxwell Lagrangian and
 generalize the framework to include also more general Yang-Mills fields. This arguments will be
treated in a forthcoming paper \cite{yangmills}.

\section{Acknowledgments}
We are grateful  to  L. Fatibene and M. Ferraris  of the
University of Torino for useful discussion on the subject. 
This work has been  partially supported  by the University of Torino
(Italy).
\pagebreak

\end{document}